\documentclass[sigconf]{acmart}

\usepackage{booktabs} 

\usepackage{graphicx}
\usepackage{graphics}
\usepackage{subfigure}
\usepackage{caption}
\usepackage{array,multirow}
\usepackage{balance}
\newcommand{\argmin}{\operatornamewithlimits{argmin}}
\graphicspath{{./figures/}}
\newcommand{\mat}[1]{{\bf #1}}
\newenvironment{sequation}{\begin{equation}\small}{\end{equation}}

\begin{document}
\fancyhead{}
\title{Understanding and Predicting Delay in Reciprocal Relations}

\author{Jundong Li}
\authornote{\label{note1}Work done when the author was an intern at Yahoo! Research}
\affiliation{%
  \institution{Arizona State University}
}
\email{jundongl@asu.edu}

\author{Jiliang Tang}
\affiliation{%
  \institution{Michigan State University}
}
\email{tangjili@msu.edu}

\author{Yilin Wang}
\authornotemark[1]
\affiliation{%
  \institution{Arizona State University}
  }
\email{ywang370@asu.edu}

\author{Yali Wan}
\authornotemark[1]
\affiliation{
  \institution{Amazon.com}
}
\email{yalwan@amazon.com}

\author{Yi Chang}
\affiliation{%
  \institution{Jilin University}
  }
\email{yichang@acm.org}

\author{Huan Liu}
\affiliation{%
  \institution{Arizona State University}
  }
\email{huan.liu@asu.edu}

\renewcommand{\shortauthors}{J. Li et al.}

\begin{abstract}
Reciprocity in directed networks points to user's willingness to return favors in building mutual interactions. High reciprocity has been widely observed in many directed social media networks such as following relations in Twitter and Tumblr. Therefore, reciprocal relations between users are often regarded as a basic mechanism to create stable social ties and play a crucial role in the formation and evolution of networks. Each reciprocity relation is formed by two parasocial links in a back-and-forth manner with a time delay. Hence, understanding the delay can help us gain better insights into the underlying mechanisms of network dynamics. Meanwhile, the accurate prediction of delay has practical implications in advancing a variety of real-world applications such as friend recommendation and marketing campaign. For example, by knowing when will users follow back, service providers can focus on the users with a potential long reciprocal delay for effective targeted marketing. This paper presents the initial investigation of the time delay in reciprocal relations. Our study is based on a large-scale directed network from Tumblr that consists of 62.8 million users and 3.1 billion user following relations with a timespan of multiple years (from 31 Oct 2007 to 24 Jul 2013).  We reveal a number of interesting patterns about the delay that motivate the development of a principled learning model to predict the delay in reciprocal relations. Experimental results on the above mentioned dynamic networks corroborate the effectiveness of the proposed delay prediction model.
\end{abstract}


\copyrightyear{2018}
\acmYear{2018}
\setcopyright{iw3c2w3}
\acmConference[WWW 2018]{The 2018 Web Conference}{April 23--27, 2018}{Lyon, France}
\acmBooktitle{WWW 2018: The 2018 Web Conference, April 23--27, 2018, Lyon, France}
\acmPrice{}
\acmDOI{10.1145/3178876.3186076}
\acmISBN{978-1-4503-5639-8/18/04}


\maketitle
\section{Introduction}

The advent and popularity of online social platforms significantly diversify the way people communicate and socialize, allowing us to share information, interact with others at a low cost and in a variety of mode. In these sites, users can form links to others. Understanding and modeling social networks and the underlying evolution mechanism have encouraged a surge of research~\cite{aggarwal2014evolutionary,aggarwal2012event,liben2007link,papadopoulos2012community,cheng2014can,li2018streaming}. Social relations can be broadly categorized into reciprocal (two-way) and parasocial relations (one-way)~\cite{horton1956mass,jun2009reciprocity}. In a directed social network like Twitter\footnote{https://twitter.com/} and Tumblr\footnote{https://www.tumblr.com/}, the social norm of reciprocity shows the tendency of a user following back to form mutual connections. Formally, it is stated as follows -- if \emph{user $u$ initiates a link to user $v$, then the action of user $v$ links back to user $u$ exhibits reciprocity}.

Reciprocal relations are widely and highly observed in directed social media networks such as following relations in Twitter~\cite{weng2010twitterrank} and Tumblr~\cite{chang2014tumblr}. They encode tighter interactions than single parasocial relations. It is evident from recent work that investigating the formation of reciprocal links and discerning the differences from parasocial links can help us gain insights on user behaviors and the dynamic mechanisms of social networks. It has implications in applications that range from friend recommendation~\cite{wang2011human}, bursty prediction~\cite{myers2014bursty}, information propagation~\cite{rodriguez2014uncovering} to viral marketing~\cite{leskovec2007dynamics}. Recent years have witnessed increasing attempts to study user reciprocal behaviors~\cite{cheng2011predicting,feng2014time,gong2013reciprocity,nguyen2010you}. Nonetheless, these methods predominantly dedicated to predicting if a parasocial link will be reciprocated back in the future given a snapshot of current network topology. As reciprocal relations stem from two parasocial relations in a back-and-forth manner, the time delay is inevitably\footnote{two parasocial relations in a reciprocal relation are not created at the same time and we refer the time gap as the time delay in this paper}and could play a non-negligible role in the formation of network dynamics and its quantification. For example, understanding the delay in reciprocal relations can help us explain how networks evolve and grow; while the length of the delay is often a good indicator to quantify the intimacy between users in the dynamic environment. In addition, the research achievements could advance many real-world applications. For instance, in friend recommender systems, our research can help determine \emph{when} we should do recommendations. Thus, it enables timely recommendation, and also, makes the recommender systems more efficient. Despite the fundamental importance, a thorough investigation on the delay in reciprocal relations is still in its infancy.

In this paper, we perform the first comprehensive investigation about the time delay in reciprocal relations. In particular, we aim to answer the following two questions -- (a) whether the delay follows certain patterns? (b) how these patterns help predict the delay in reciprocal relations automatically? For question (a), we study some potential factors that may influence the delay from the temporal and structural perspectives and our major understandings are summarized as follows:
\begin{itemize}
\item From the temporal perspective, (1) the delay is affected by the time when users join the social network and how long they have been in the network when the reciprocal relations are initiated; (2) the delay presents weekly patterns -- the delay is shorter when reciprocal relations originate on weekends; (3) the delays of two consecutive reciprocal relations from the same user are not necessarily related, i.e., the previous immediate delay is not a good estimator of the current one from the same user.
\item From the structural perspective, (1) the delay is related to indegrees and outdegrees of users -- users with higher indegrees tend to get reciprocated faster but reciprocates back to others slower; and outdegrees present similar patterns as indegrees; (2) the delay between users that share a large number of common followees and followers are often shorter than those without much commonalities.
\end{itemize}

\noindent To answer the question (b), we propose a framework that can predict the delay in reciprocal relations automatically. Empirical results on the large-scale Tumblr network demonstrate that the proposed model can accurately predict the delay in reciprocal relations.

The rest of paper is organized as follows. In Section 2, we present a large-scale dynamic social network to study and show its basic statistics. Section 3 presents a comprehensive study on the time delay in reciprocal relations. In Section 4, we introduce the proposed model to predict the delay automatically. Section 5 presents experimental results with discussions. Section 6 introduces related work and Section 7 concludes the paper with future research directions.

\section{Data}

We perform our study on a microblogging platform - Tumblr. In Tumblr, users can follow the blogs of others without necessarily following back, which naturally forms a directed social network. By Feb 2016, there are around 550 million monthly active users and 280.4 million blog posts in Tumblr. In this study, we collect a dynamic network that consists of all user following relationships from 31 Oct 2007 to 24 Jul 2013 (totally 2,094 days)~\cite{chang2014tumblr}. By its end date, the network contains 62.8 million users and 3.1 billion edges. In this study, we specify the time granularity as one day and use $t=0$ and $t=2093$ to denote the start date (10/31/2007) and end date (07/24/2013), respectively. The detailed statistics of the dataset are listed in Table~\ref{table:datasets}.
\begin{table}
\centering
\begin{tabular}{c|c} \hline
Description& Tumblr \\ \hline \hline
$\#$ of Time Stamps (days) & 2,094 \\ \hline
$\#$ of Users at $t = 0$ & 13,864  \\ \hline
$\#$ of Links at $t = 0$ & 51,741 \\ \hline
$\#$ of Users at $t = 2,093$ & 62,848,996 \\ \hline
$\#$ of Links at $t = 2,093$ &  3,132,353,040 \\ \hline
\end{tabular}
\caption{Detailed statistics of the Tumblr dataset.}
\label{table:datasets}
\end{table}

\begin{figure}[!t]
\centering
\begin{minipage}{0.235\textwidth}
\centering
\subfigure[$\#$ of nodes\label{fig:numnodes}]
{\includegraphics[width=\textwidth]{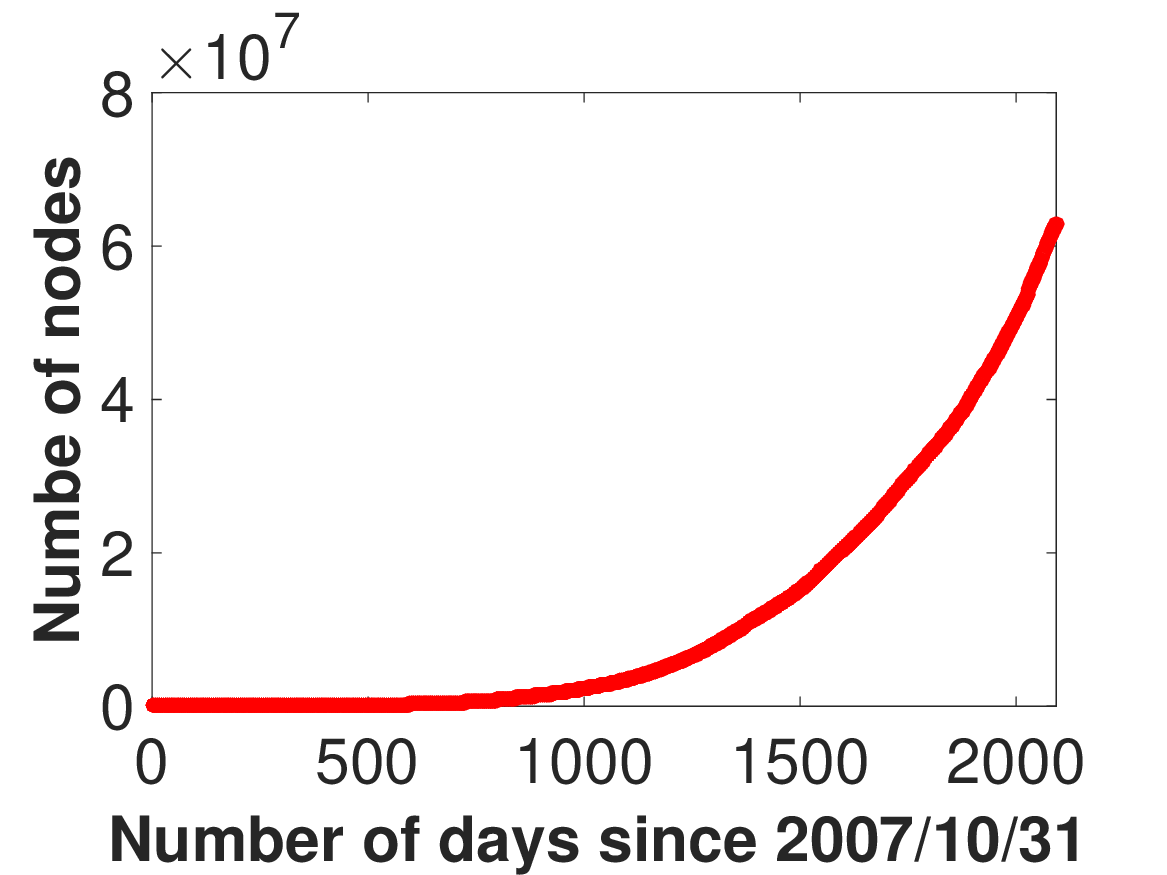}}
\end{minipage}
\begin{minipage}{0.235\textwidth}
\centering
\subfigure[$\#$ of links\label{fig:numedges}]
{\includegraphics[width=\textwidth]{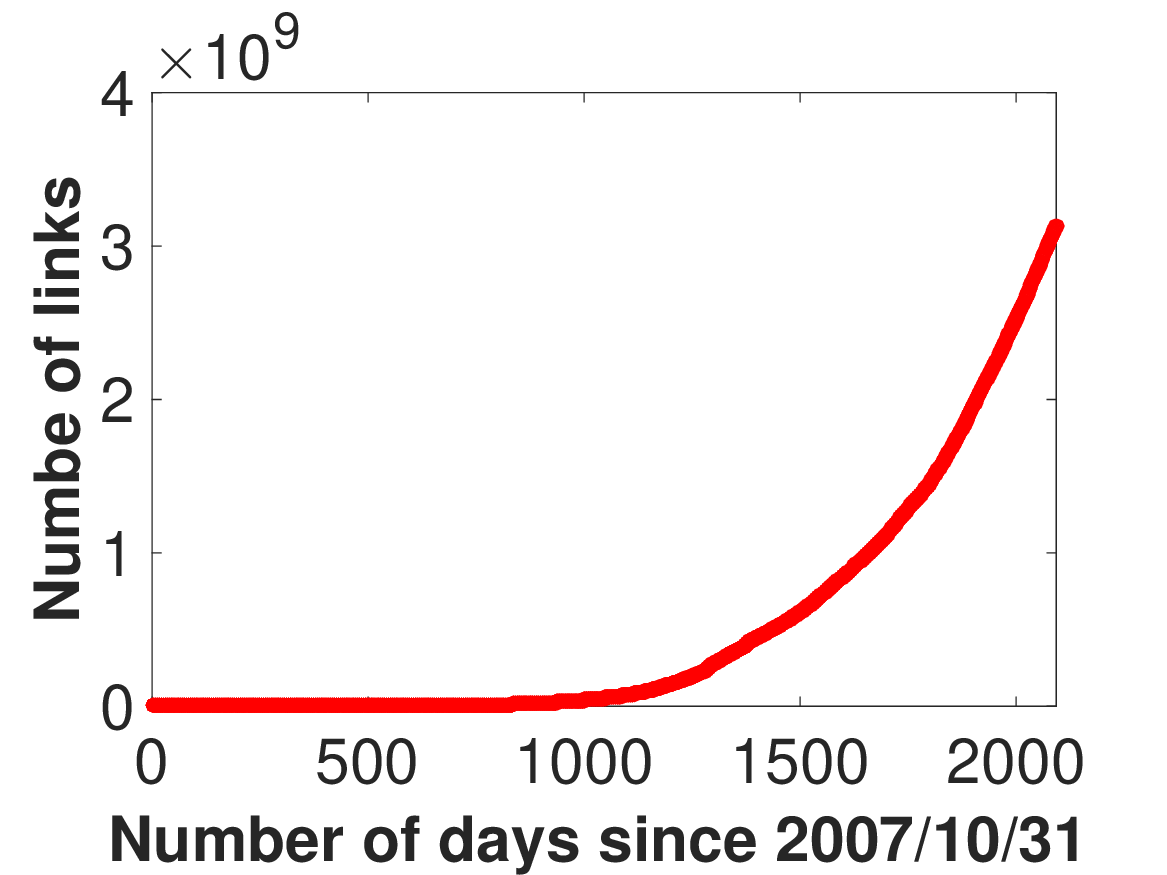}}
\end{minipage}
\begin{minipage}{0.235\textwidth}
\centering
\subfigure[$\#$ of reciprocities\label{fig:numreciprocity}]
{\includegraphics[width=\textwidth]{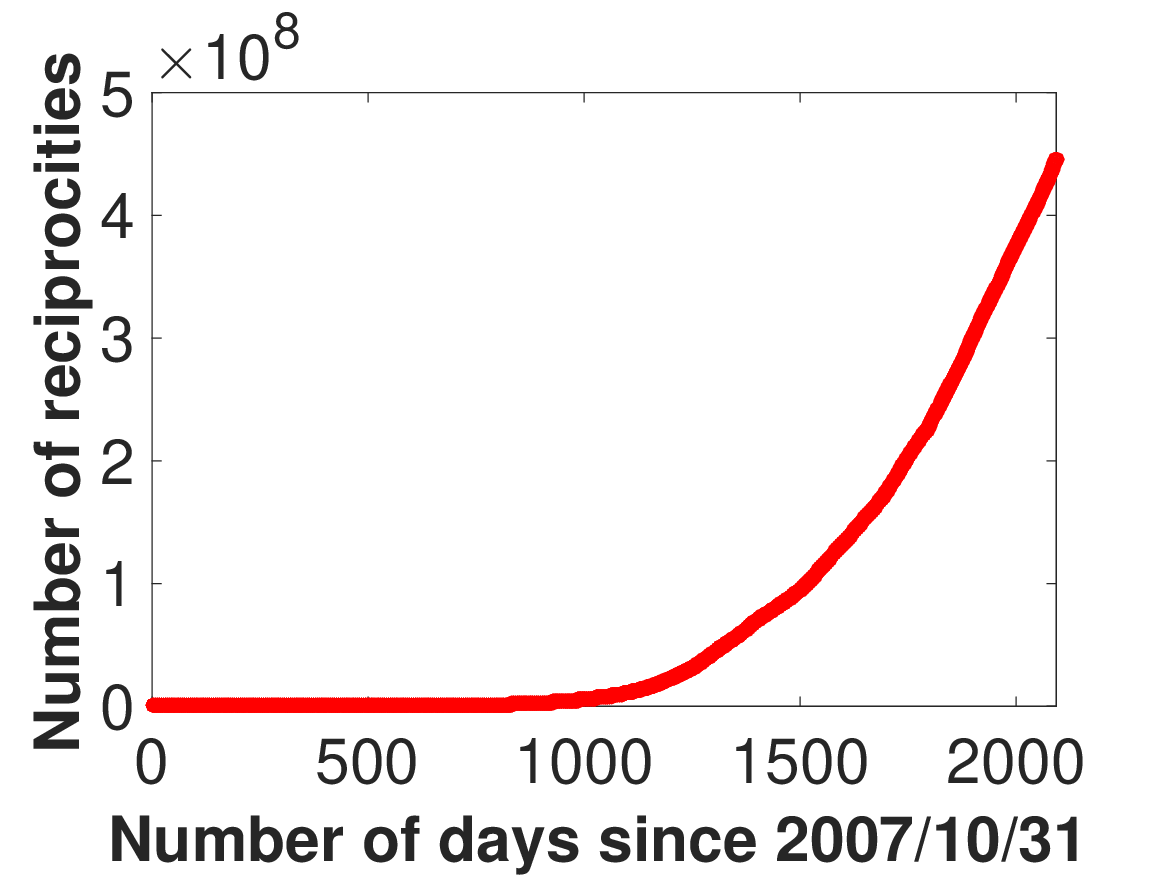}}
\end{minipage}
\begin{minipage}{0.235\textwidth}
\centering
\subfigure[reciprocity rate\label{fig:reciprocity}]
{\includegraphics[width=\textwidth]{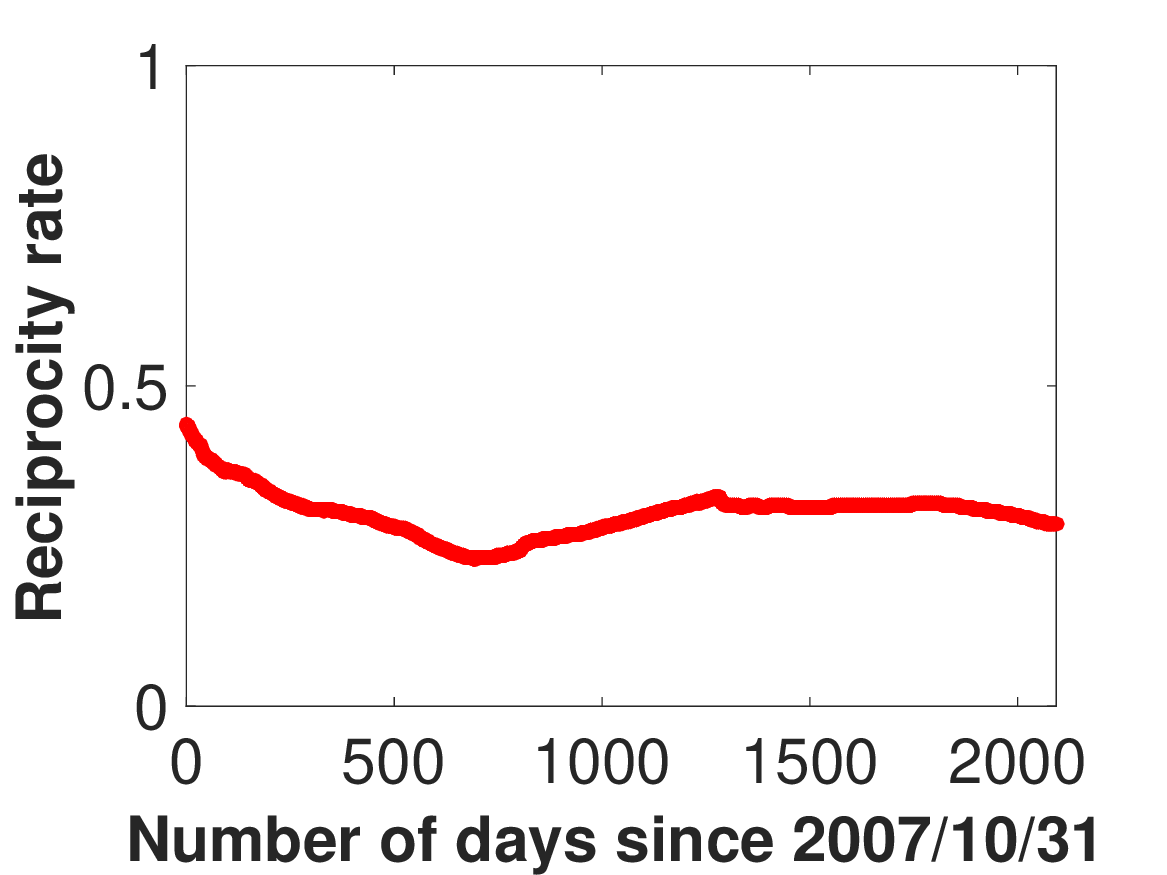}}
\end{minipage}
\caption{Evolution of network statistics over time.}
\label{fig:tumblr_properties}
\end{figure}

\begin{figure}[!t]
\centering
\begin{minipage}{0.235\textwidth}
\centering
\subfigure[average degree\label{fig:avgdegree}]
{\includegraphics[width=\textwidth]{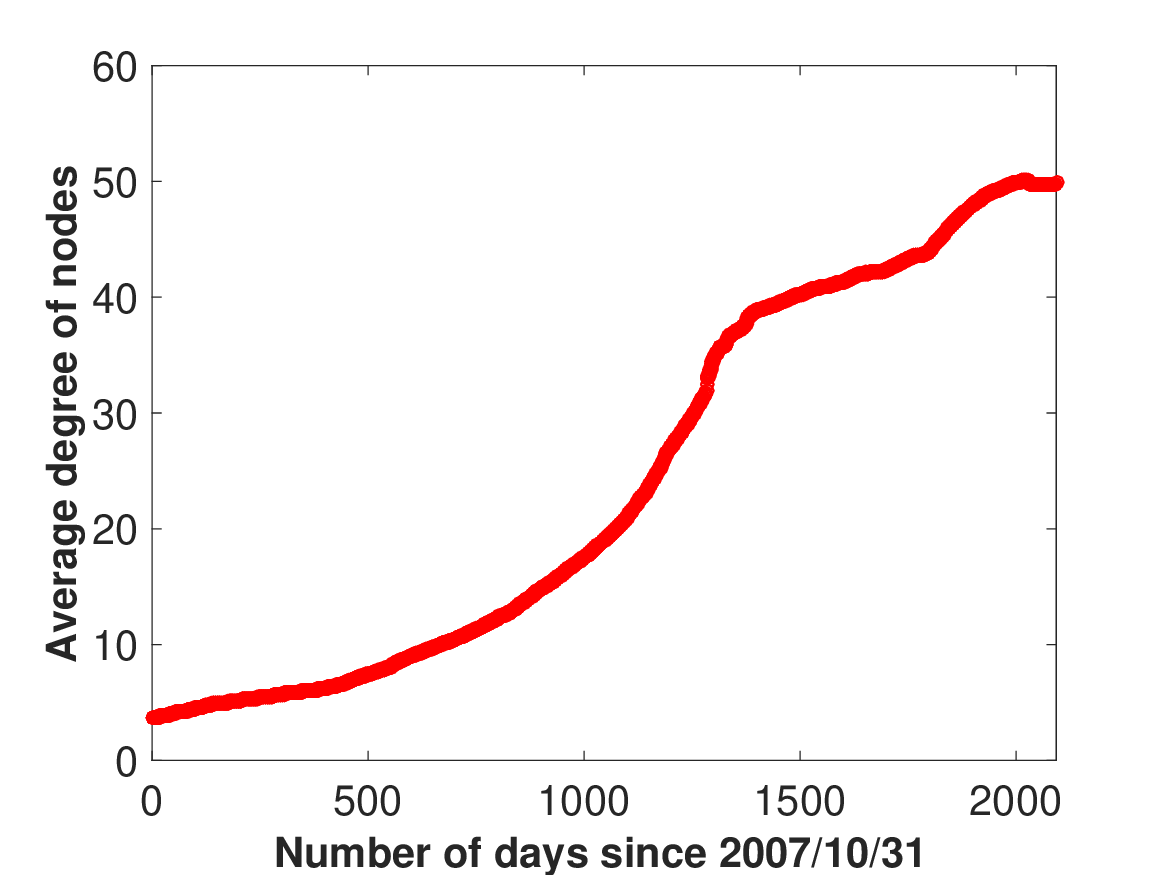}}
\end{minipage}
\begin{minipage}{0.235\textwidth}
\centering
\subfigure[$e(t)$ versus $n(t)$ \label{fig:numnodesnumedges}]
{\includegraphics[width=\textwidth]{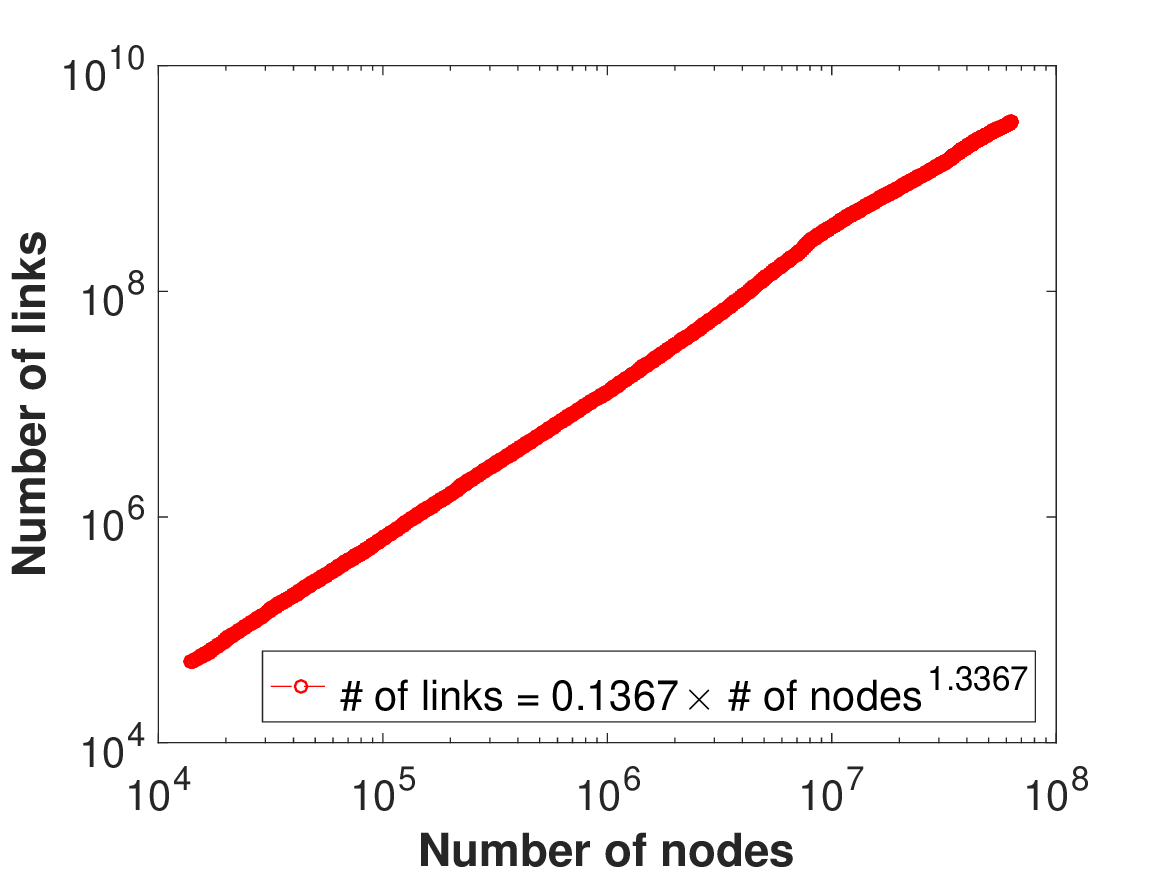}}
\end{minipage}
\caption{Network densifies over time.}
\label{fig:densify}
\end{figure}

{\it Evolution patterns of the Tumblr network:} We first examine whether the directed Tumblr network presents principled evolution patterns that have been observed on other dynamic social networks.  Figure~\ref{fig:numnodes}, Figure~\ref{fig:numedges}, Figure~\ref{fig:numreciprocity} display the growth of nodes, links, reciprocal relations, respectively. The network growth is relatively slower at the very beginning. From the year of 2010, the Tumblr network experienced an exponential growth. Similar observations are also reported on other social media sites\footnote{http://dstevenwhite.com/2011/12/29/social-media-growth-2006-2011/}. Meanwhile, we show how the reciprocity of the Tumblr network changes over time in Figure~\ref{fig:reciprocity}. It shows that the reciprocity is relatively high right before the network exponential growth. Afterwards, the reciprocity rate gradually declines and then remains rather stable, which is consistent with the findings in~\cite{nguyen2010you}. In addition, network densification~\cite{leskovec2005graphs} is a widely perceived pattern of dynamic networks. It suggests that networks are gradually densified. To examine this property, we first empirically show how the average indegree (outdegree) of the network nodes changes in Figure~\ref{fig:avgdegree}. It is clear that the average degree increases over time, thus the network is becoming denser. To further corroborate this observation, we also show the number of links at each time stamp ($e(t)$) versus the number of nodes at each time step ($n(t)$) in Figure~\ref{fig:numnodesnumedges} in a log-log scale. We observe that the number of links indeed increases superlinearly w.r.t. number of nodes with a slope of $a=1.3367$. The observations are very consistent with those in~\cite{leskovec2005graphs}. Given the scale of the network and the aforementioned observations, it is safe to conclude that the Tumblr network is a typical and representative directed social network. Therefore we can rely on it to study the time delay in reciprocal relations.

\begin{figure}[!t]
\centering
\includegraphics[width=0.45\textwidth]{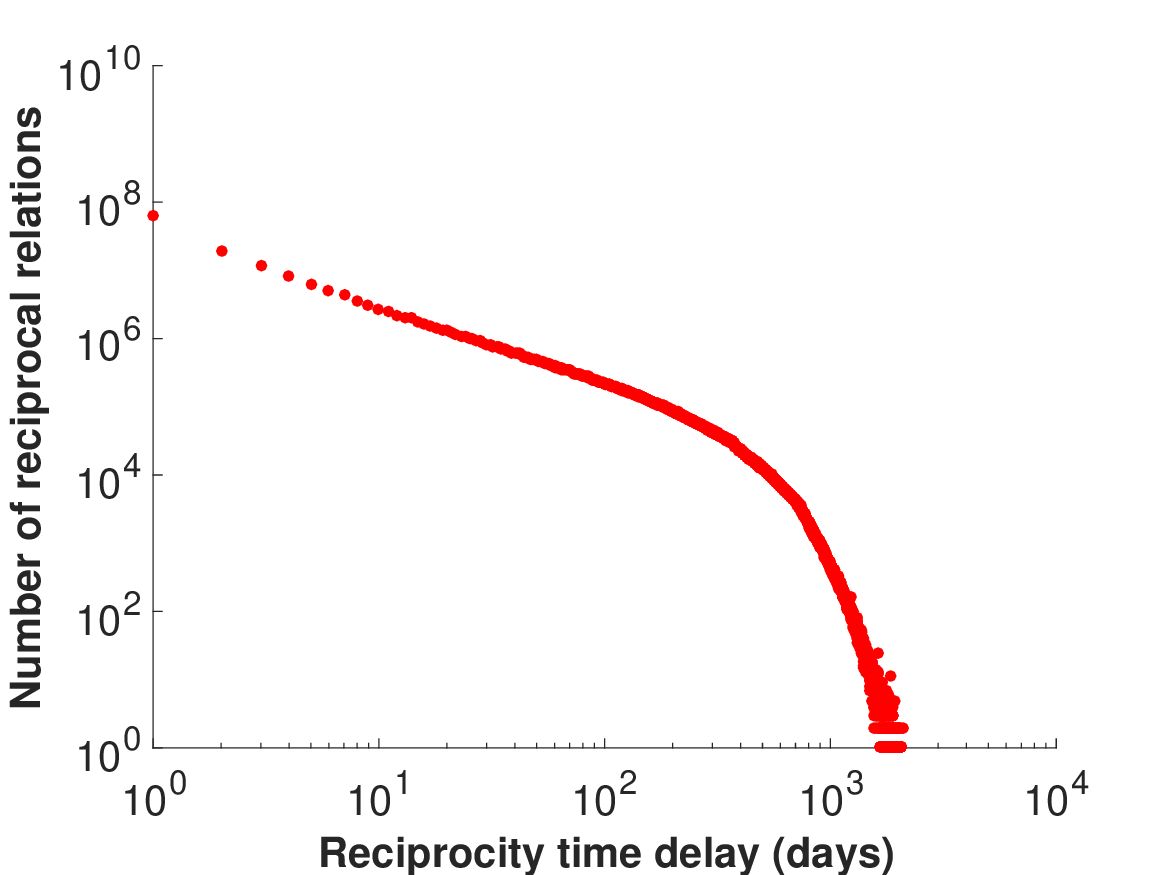}
\caption{Reciprocity time delay distribution.}
\vspace{-0.2in}
\label{fig:reciprocity_duration_distribution}
\end{figure}

{\it Distribution of delay in reciprocal relations:} We first formally define the concept of \emph{the time delay} in reciprocal relations. For a reciprocal relation with two users $u$ and $v$, assume that user $u$ initiates a parasocial relation to user $v$ at time stamp $t_{1}$ and user $v$ follows back to user $u$ at time stamp $t_{2}$ ($t_{2}\geq t_{1}$). Then the delay in the reciprocal relation is defined as the difference between $t_{2}$ and $t_{1}$, i.e., $t_{2}-t_{1}$. With the definition, we plot its distribution until $t=2093$ for all {\it 446,116,001} reciprocal relations in Figure~\ref{fig:reciprocity_duration_distribution}. We first notice that the delay follows a power law distribution with an exponential cutoff~\cite{newman2005power}. It means that the power law rule holds for short delays and then it declines exponentially for long ones. The reason is that for some delays in reciprocal relations whose initiate time is close to the end date of the dataset, their reciprocal behaviors have been not fully observed yet. This phenomenon can also be explained by right censoring issue in survival analysis~\cite{wu1988estimation}. In a nutshell, we observe that the time delay exists in a large portion of reciprocal relations that suggests the necessity of the studied problem

\section{Analysis on the Time Delay in Reciprocal Relations}
To reveal the delay patterns, we focus on understanding the factors that can influence the length of the delay. Next, we carry out a comprehensive study of the time delay from the temporal and structural perspectives. To ease the following presentations, we now formally define each delay as a tuple $<u,v,t_{1},t_{2},t_{2}-t_{1}>$. It denotes that user $u$ (referred as the source user) follows user $v$ (referred as the target user) at the time stamp $t_{1}$, and as a favor, user $v$ follows back to user $u$ at the time stamp $t_{2}$, the delay of the reciprocal relation is $t_{2}-t_{1}$, where $t_{1}\leq t_{2}$.

\subsection{Temporal Analysis}
\begin{figure}[!t]
\centering
\begin{minipage}{0.235\textwidth}
\centering
\subfigure[Source user\label{fig:src_join_time_duration_avg}]
{\includegraphics[width=\textwidth]{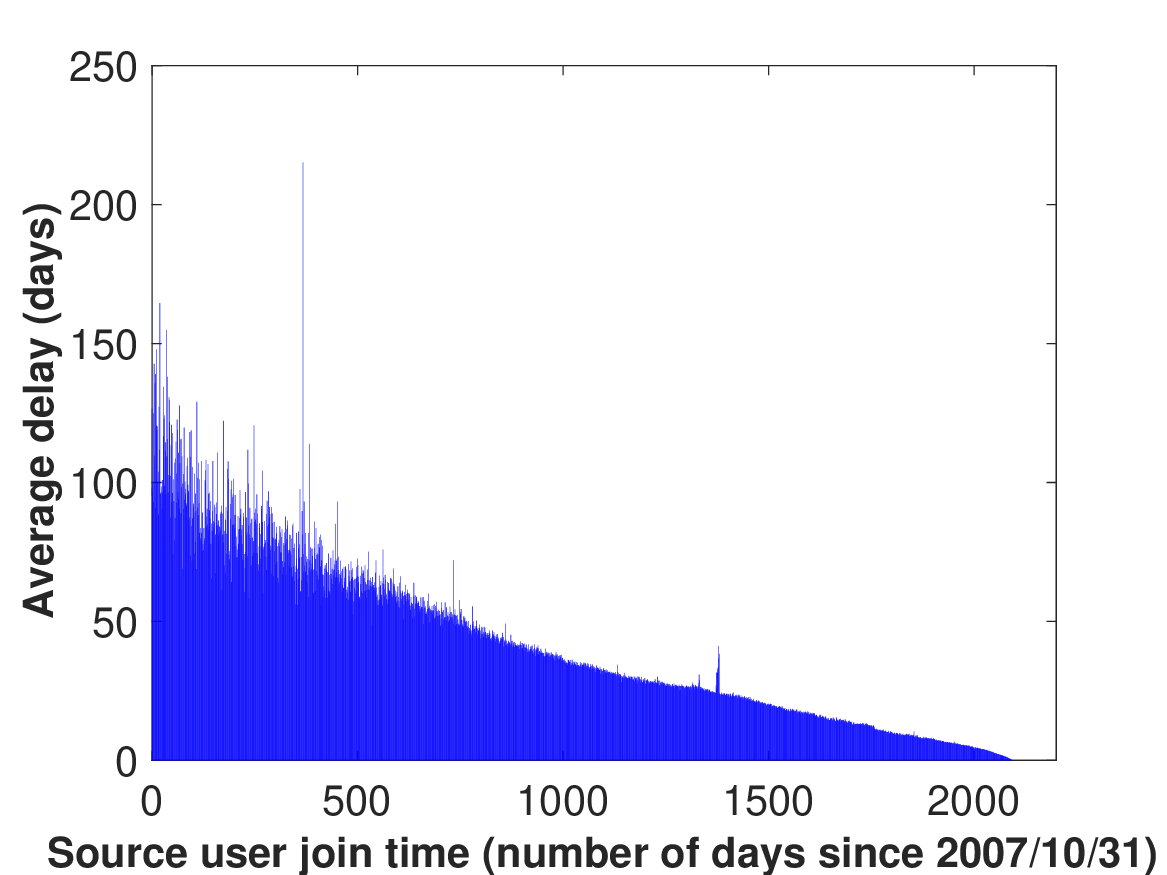}}
\end{minipage}
\begin{minipage}{0.235\textwidth}
\centering
\subfigure[Target user\label{fig:des_join_time_duration_avg}]
{\includegraphics[width=\textwidth]{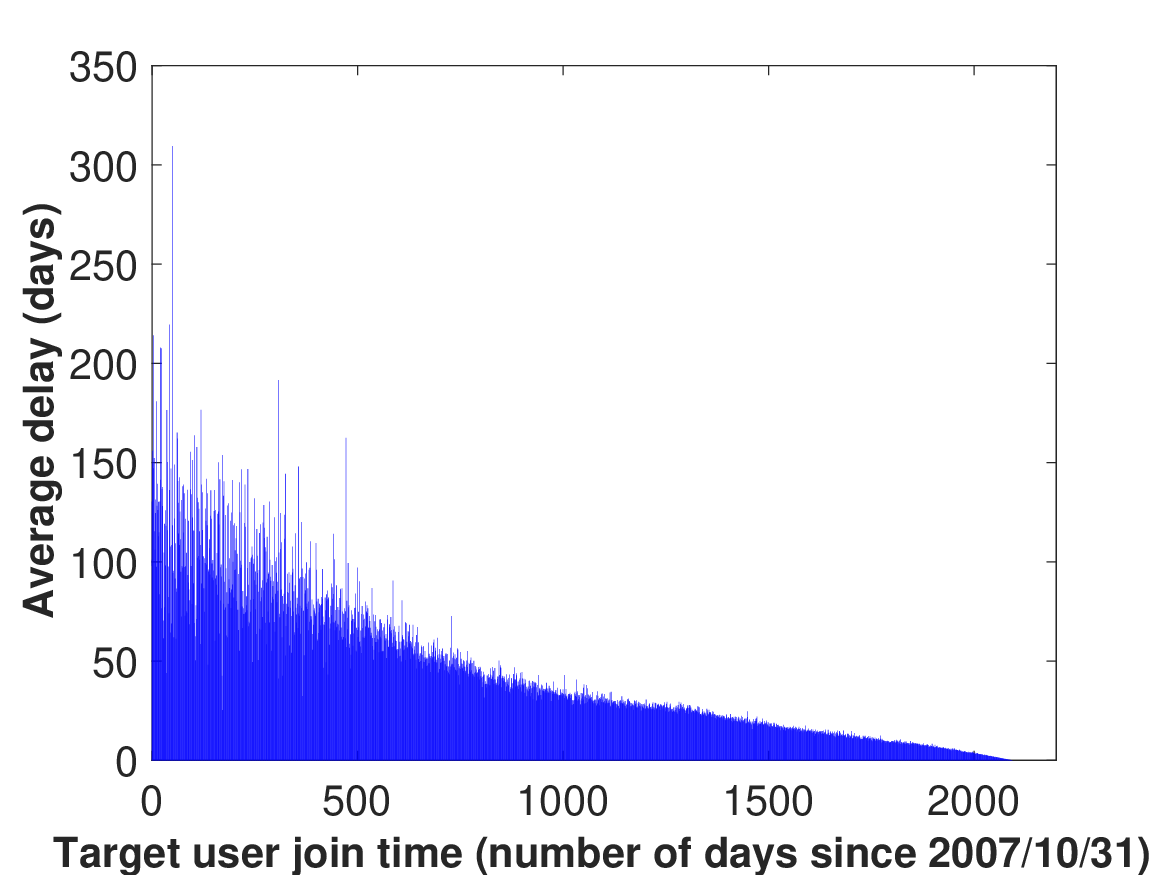}}
\end{minipage}
\caption{Average delay versus when users join the network.}
\label{fig:join_time_duration_avg}
\end{figure}

{\it Does the user joining time matter?} Our first analytical task is to check if the delay is related to the time when users join the network. Assume that the joining time of the reciprocal source user $u$ and the reciprocal target user $v$ are $t_{u}$ and $t_{v}$ respectively, a natural question is: does the user joining time affect the delay? To answer this question, we plot the average delay of reciprocal relations with respect to the user joining time stamps in Figure~\ref{fig:join_time_duration_avg}. In general, the average delay gradually decreases with the shifting of the user joining time. In particular, we observe that (1) for source users who join the network earlier, the target users need more time to follow them back; and (2) for target users who join the network earlier, the delay is longer. Intuitively, users gradually become less active in terms of establishing social relations; as a result, the delays of reciprocal relations of earlier joined users become longer.


\begin{figure}[!t]
\centering
\begin{minipage}{0.235\textwidth}
\centering
\subfigure[Average delay versus reciprocal initiating day of the week\label{fig:week}]
{\includegraphics[width=\textwidth]{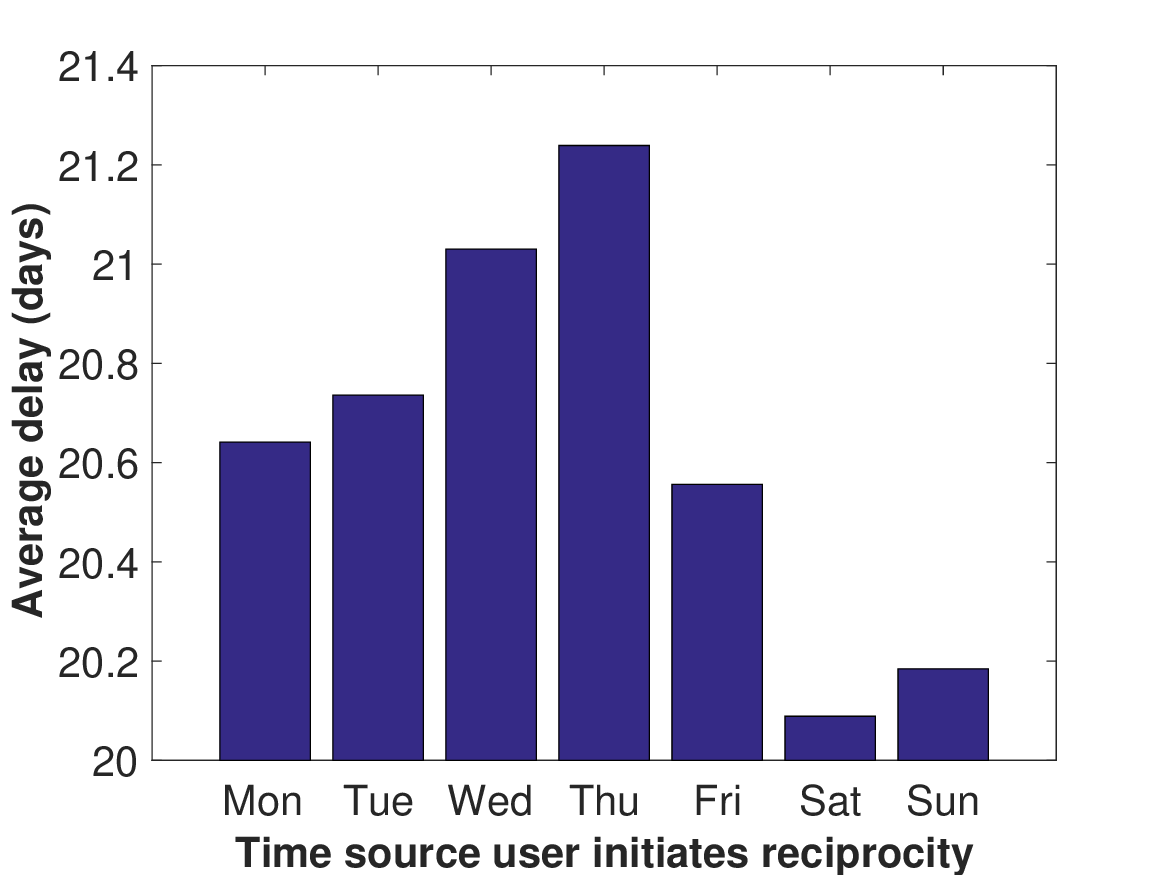}}
\end{minipage}
\begin{minipage}{0.235\textwidth}
\centering
\subfigure[$\#$ reciprocities versus reciprocal completing day of the week \label{fig:week2}]
{\includegraphics[width=\textwidth]{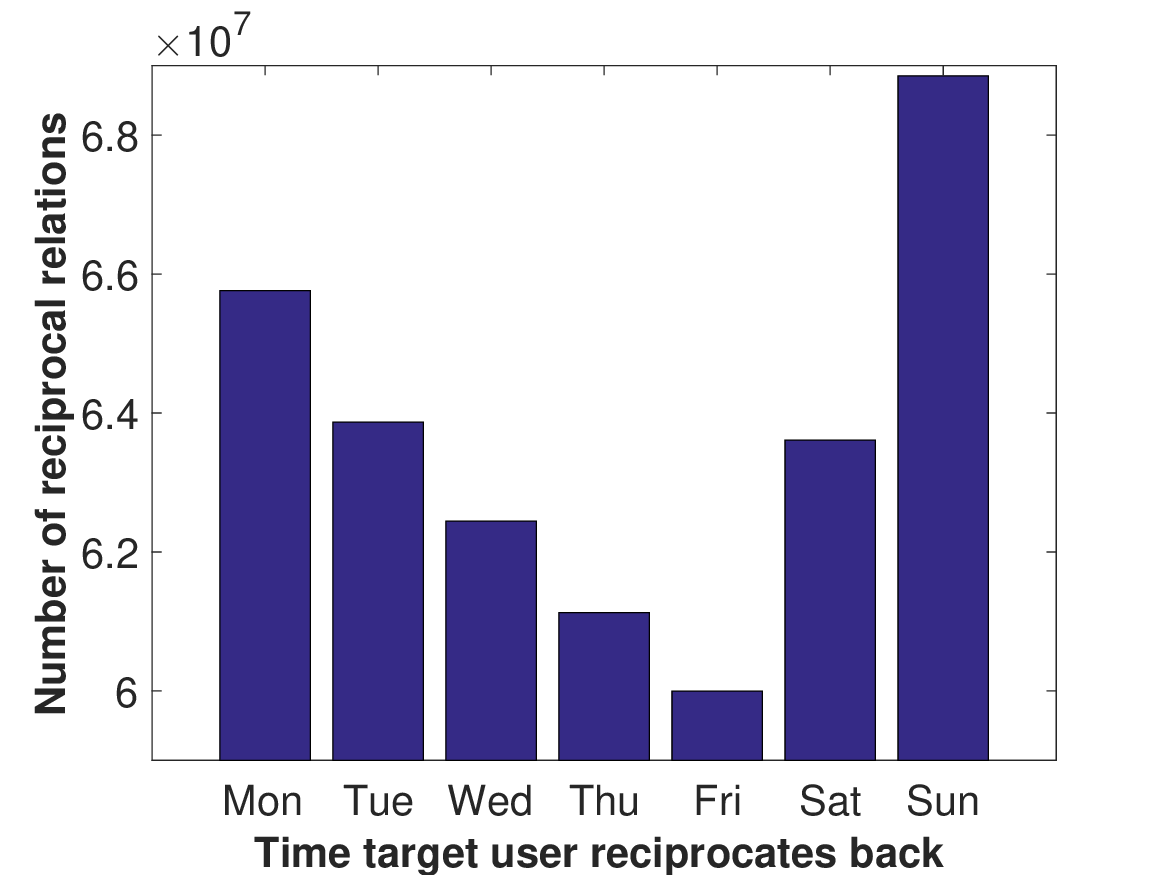}}
\end{minipage}
\caption{Weekly delay patterns.}
\label{fig:weekpattern}
\end{figure}

{\it Weekly delay patterns:} Weekly patterns have been observed in various types of user behaviors such as check-in behaviors~\cite{chang2011location} that motivate us to investigate the weekly delay patterns in reciprocal relations. The average delay with respect to the day of the week is depicted in Figure~\ref{fig:week}. It can be observed that the delay of the reciprocal relations that are initiated on Saturday and Sunday is shorter; while the average delay on Thursday is longest. To understand these patterns, we also calculate the number of reciprocal relations completed each day in a week and the results are shown in Figure~\ref{fig:week2}. On average, there are much more reciprocal relations completed on weekends than weekdays so the reciprocal relations initiated on weekends are likely to be finished within $0$ or $1$ day. This observation is helpful to explain why the delay in weekends is shorter.  Furthermore, from Monday to Thursday, the number decreases while it increases from Friday to Sunday. This finding can help us understand why the delay on Thursday is the longest.

\begin{figure}[!t]
\centering
\begin{minipage}{0.235\textwidth}
\centering
\subfigure[MAE\label{fig:previousMAE}]
{\includegraphics[width=\textwidth]{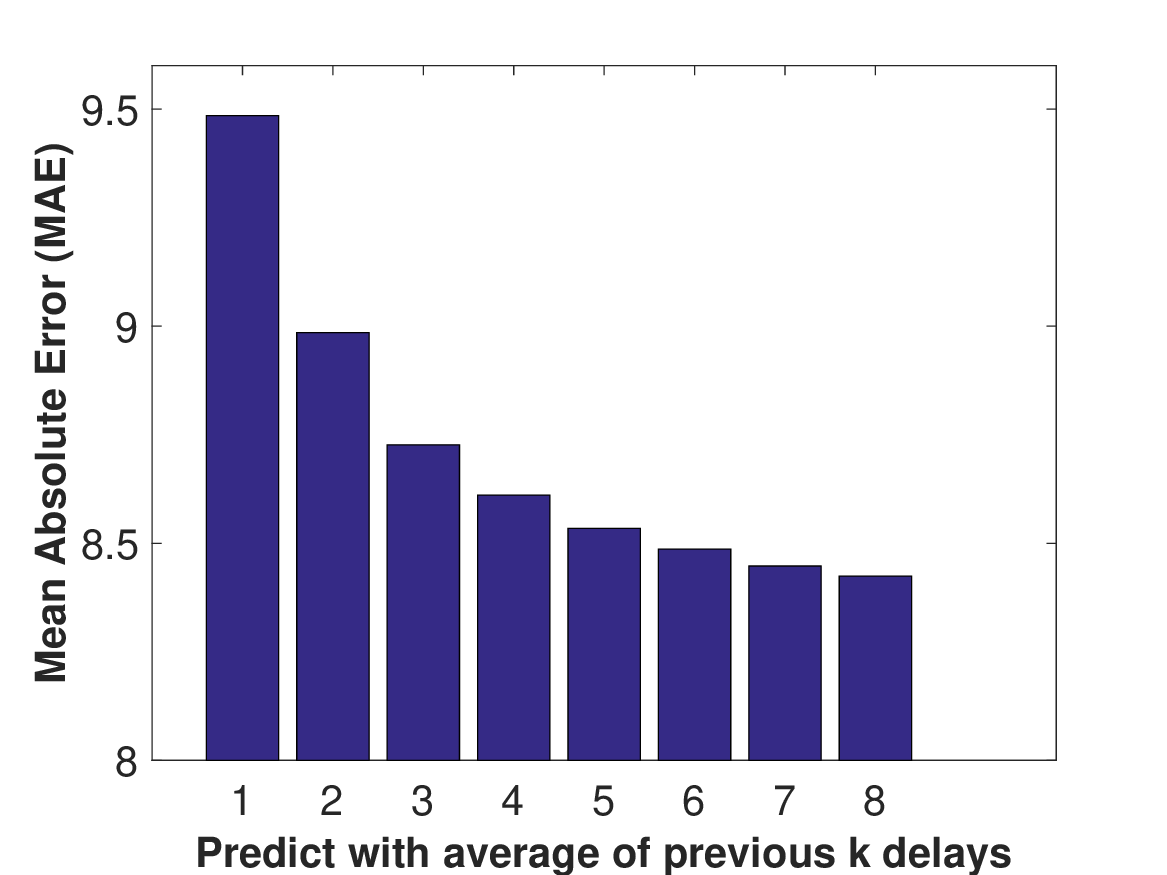}}
\end{minipage}
\begin{minipage}{0.235\textwidth}
\centering
\subfigure[RMSE\label{fig:previousRMSE}]
{\includegraphics[width=\textwidth]{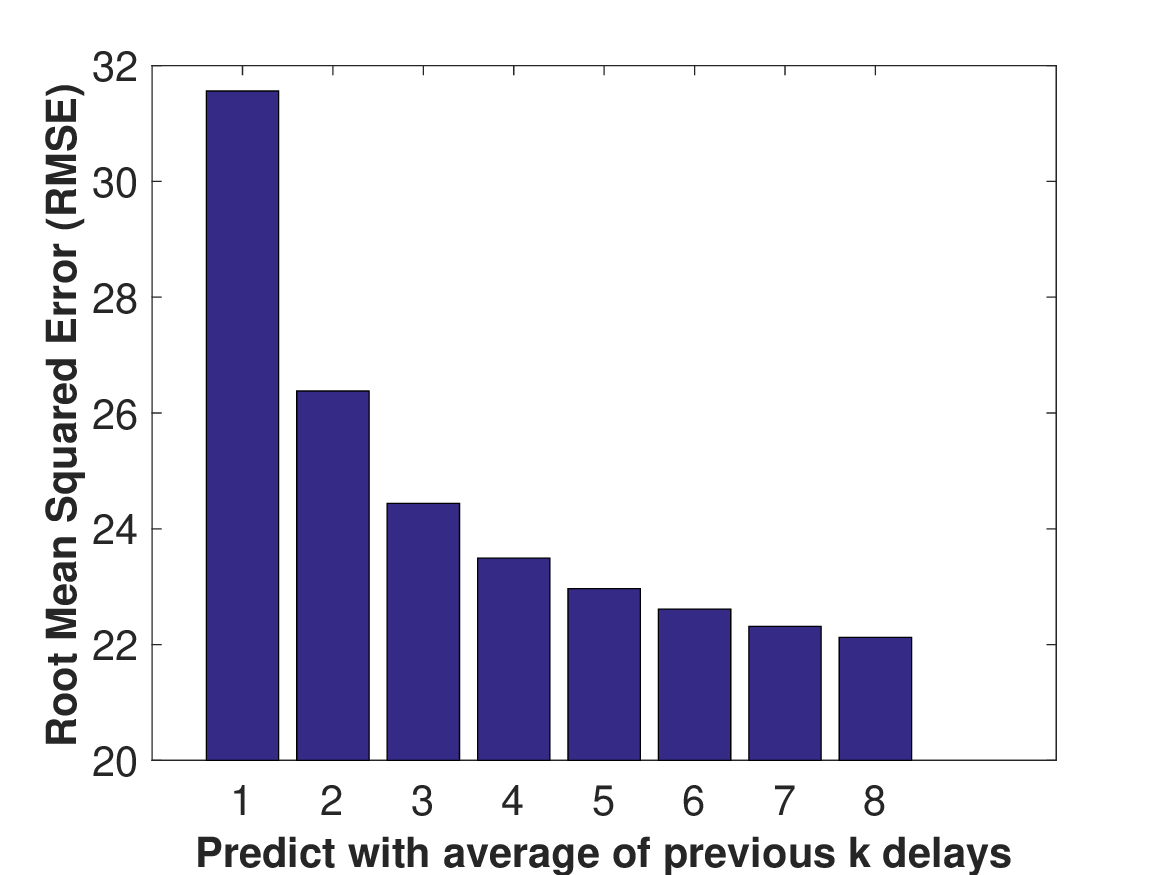}}
\end{minipage}
\caption{Delay prediction performance using the average of previous $k$ delays.}
\label{fig:previous}
\end{figure}

{\it Sequential patterns:} We can consider the following back behaviors in the timeline of one user as a sequence and sequential patterns have been widely observed and studied in temporal user behavior analysis. For example, users' check-in behaviors in LBSNs are related to their previous check-ins especially the immediate previous one~\cite{gao2012exploring}. To study the sequential patterns of users' reciprocal behaviors, for each user we investigate to which extend the current delay can be predicted by the average of his/her previous $k$ delays. We vary the value of $k$ from $1$ to $8$ with a step of $1$. To alleviate the biases from the extraordinarily long delays, we filter the reciprocal relations that are delayed exceeding 50 days as more than 90\% reciprocal relations completes within 50 days. The prediction performance are evaluated by two widely used metrics, mean absolute error (MAE) and root mean squared error (RMSE). They are defined as:
\begin{equation}
\begin{split}
MAE &=\frac{1}{n}\sum_{i=1}^{n}|d_{i}-\hat{d_{i}}| \\
RMSE &= \sqrt{\frac{\sum_{i=1}^{n}(d_{i}-\hat{d_{i}})^{2}}{n}},
\end{split}
\end{equation}
where $d_{i}$ and $\hat{d_{i}}$ denote true delay and the predicted delay, respectively. The prediction results are shown in Figure~\ref{fig:previous}. We can observe that if we only use the delay of the immediately previous one (or $k = 1$) to make the prediction, the prediction error is the highest that is different from observations about other user behaviors such as check-in behaviors~\cite{gao2012exploring}. When the value of $k$ gradually increases, the prediction error tends to decrease accordingly. This observation suggests that the average delay in a certain period is a better indicator than the immediately previous one. In other words, the strong correlation between the current behavior and the immediate previous behavior does not necessarily hold for delays in reciprocal relations. To further probe the reason of this phenomena, we investigate all reciprocal relations sequentially and find out that: (1) a consideration portion of users tend to reciprocate back to their recent and earlier followers at the same time; and (2) users do not necessarily follow the initiating orders to complete reciprocal relations (e.g., users could delay earlier initiating reciprocal relations; while finishing recent ones).

\subsection{Network Structural Analysis}

In a directed network,  in and out degrees are two importance characteristics of users and structural similarity is one of the most important characteristics for pairs of users. In this subsection, we investigate the impact of these importance characteristics on the time delay in the reciprocal relations.

{\it The impact of in and out degrees:} Celebrities in social media usually attract more followers than normal users and active users inclined to follow more users. Hence, it is natural to question -- is the delay related to in and out degrees of users? The indegrees of users at the initiating time of reciprocal relations can indicate their social statuses; while the outdegrees of users can reflect their user activities. To differentiate users with different indegrees, we categorize users with more than 2000 followers as high indegrees, users with less than 10 followers as low indegrees and the remaining users as normal. Similarly, in order to discriminate users with different outdegrees, we consider users who follow more than 2000 users as high outdegrees, users with less than 10 following relationships as low outdegrees, and the other users as normal users.
Detailed categorization of users according to their social status and activity is illustrated in Table~\ref{table:inoutdegree}.
\begin{table}
\centering
\begin{tabular}{c|c|c|c} \hline
& <10 & 10 to 2000 & >=2000\\ \hline \hline
outdegree & low & normal & high \\ \hline
indegree & low & normal & high  \\ \hline
\end{tabular}
\caption{Categorization of Users by In and Out Degree.}
\label{table:inoutdegree}
\end{table}

\begin{figure}[!t]
\centering
\begin{minipage}{0.235\textwidth}
\centering
\subfigure[Source user\label{fig:src_indegree}]
{\includegraphics[width=\textwidth]{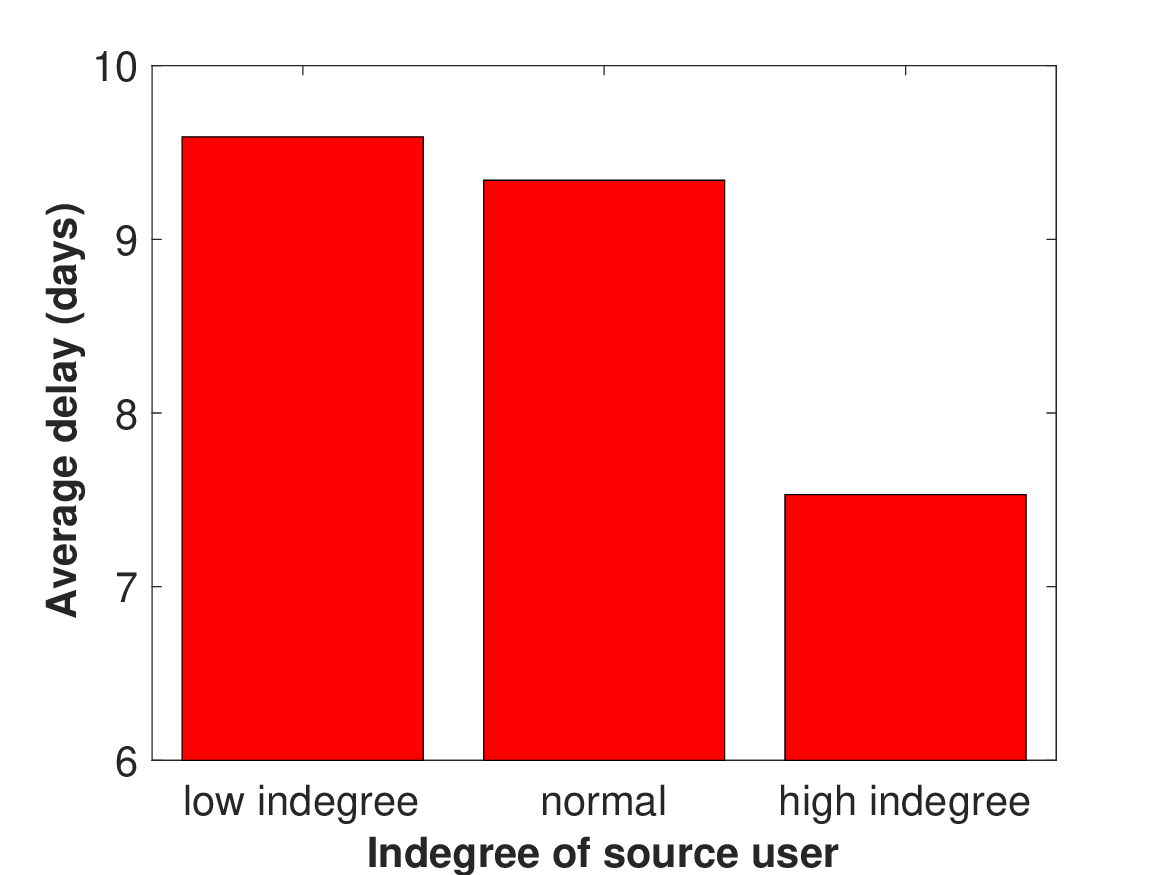}}
\end{minipage}
\begin{minipage}{0.235\textwidth}
\centering
\subfigure[Target user\label{fig:des_indegree}]
{\includegraphics[width=\textwidth]{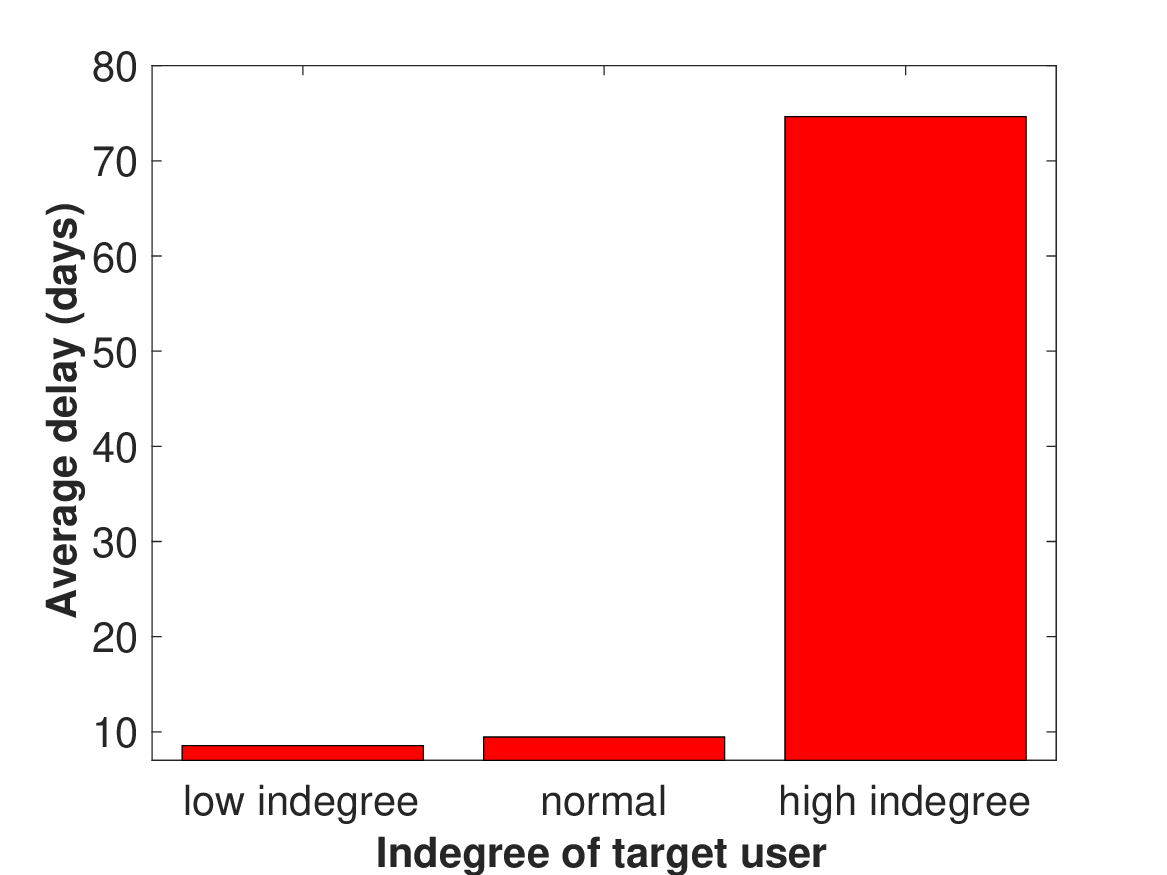}}
\end{minipage}
\caption{Average delay versus user indegrees (social status) at the initiating time of reciprocal relations.}
\label{fig:src}
\end{figure}

Firstly, we show how users of different indegrees impact the reciprocal delays. As each reciprocal relation involves two users, we show the differences of the average delay of the source and target users at the initiating time in Figure~\ref{fig:src_indegree} and Figure~\ref{fig:des_indegree}, respectively. It can be observed from Figure~\ref{fig:src_indegree} that as indegrees of source users increases, the average delay becomes shorter. It suggests that the following relations initiated by famous users (i.e., celebrities) get reciprocated faster than those of low indegrees. Another interesting finding from Figure~\ref{fig:des_indegree} is that high indegree users normally take longer (i.e., 7 times in our studied dataset) to follow back to their followers than others. Normal users spend relatively a little bit longer time (around 0.9 days) to reciprocates back to their followers than low indegree users. In summary, as the source users, higher indegrees indicate shorter delay; while as the target users, higher indegrees mean longer delay.

\begin{figure}[!t]
\centering
\begin{minipage}{0.235\textwidth}
\centering
\subfigure[Source user\label{fig:src_outdegree}]
{\includegraphics[width=\textwidth]{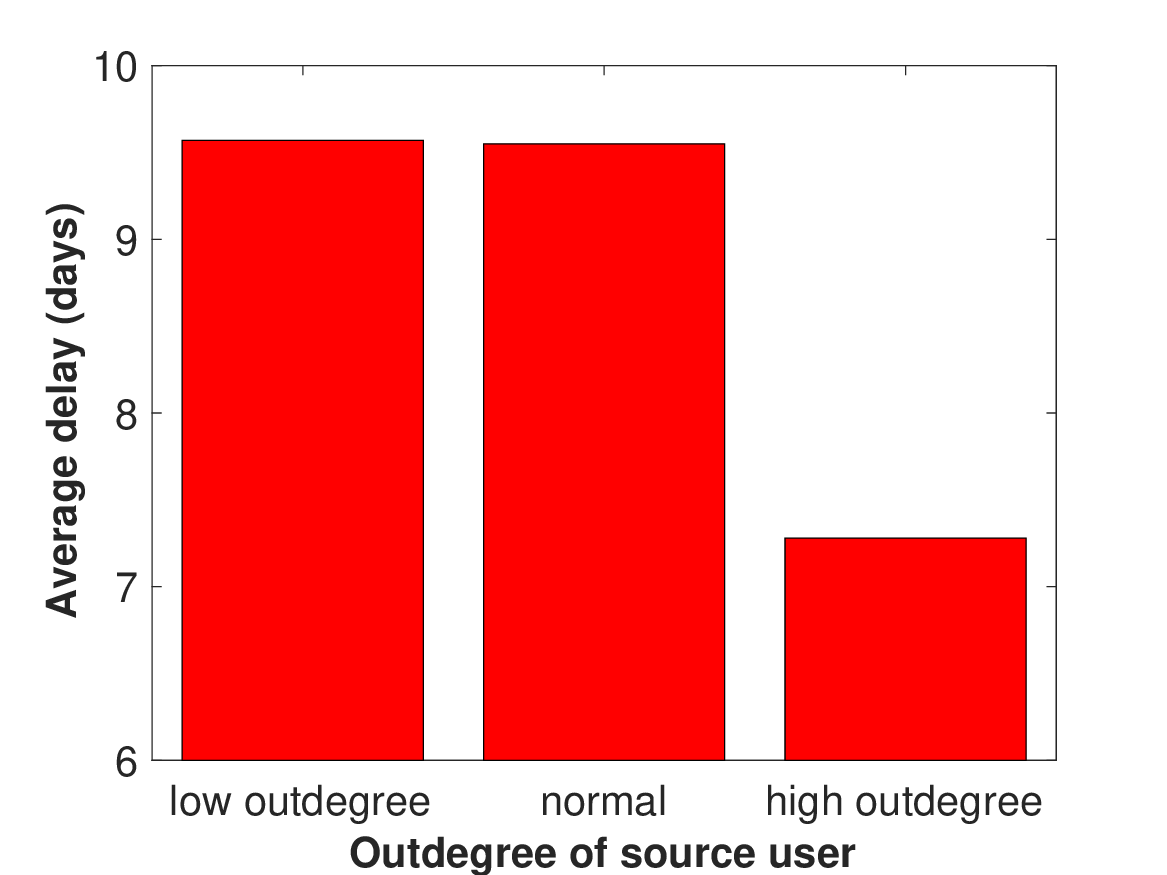}}
\end{minipage}
\begin{minipage}{0.235\textwidth}
\centering
\subfigure[Target user\label{fig:des_outdegree}]
{\includegraphics[width=\textwidth]{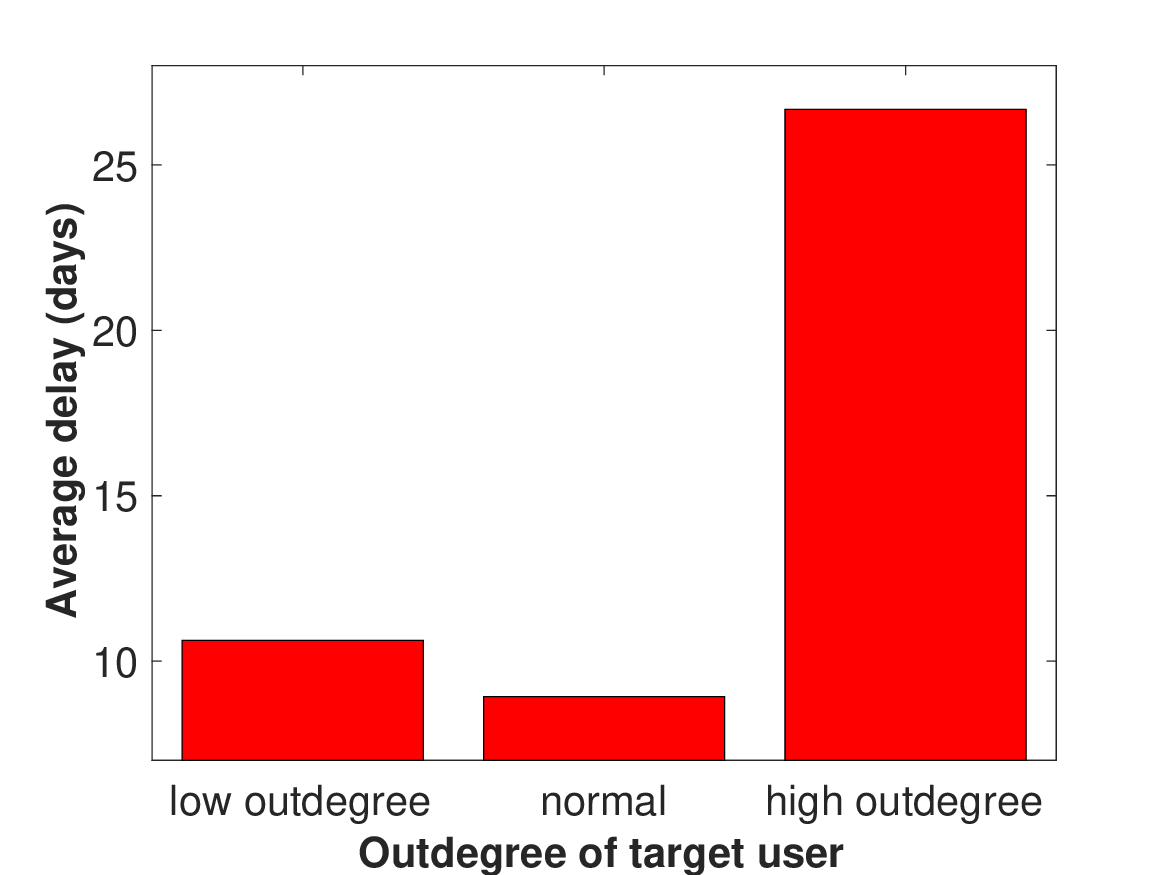}}
\end{minipage}
\caption{Average delay versus user outdegrees (user activity) at the initiating time of reciprocal relations.}
\label{fig:des}
\end{figure}

Second, we examine how the delay is related to the user outdegrees. We present the average delay (source and target users) with different outdegrees in Figure~\ref{fig:des}. Figure~\ref{fig:src_outdegree} shows that high outdegree users are likely to get their followers reciprocated back faster than low outdegree users. High outdegrees often mean more activities and higher visibility that could shorten the delay of reciprocal relations. In contrast, high outdegree users delay longer to follow back to their followers. In a typical directed social network, users' indegrees and outdegrees are almost balanced -- high indegrees often mean high outdegrees~\cite{szell2010multirelational}. That can support the reason why we make very similar observations about the impacts of in and out degrees on the time delay in reciprocal relations.

\begin{figure}[!t]
\centering
\begin{minipage}{0.235\textwidth}
\centering
\subfigure[$\#$ common followees\label{fig:cofollowing}]
{\includegraphics[width=\textwidth]{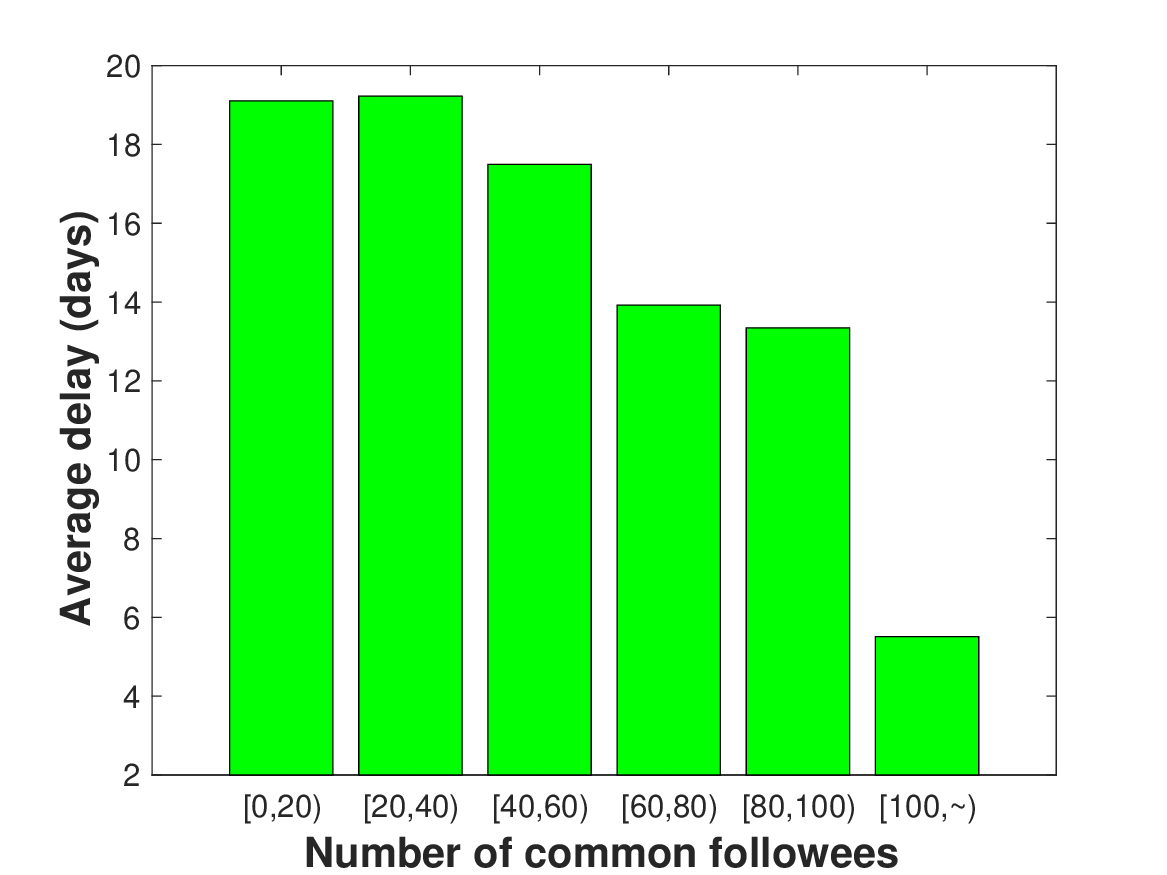}}
\end{minipage}
\begin{minipage}{0.235\textwidth}
\centering
\subfigure[$\#$ common followers\label{fig:cofollowee}]
{\includegraphics[width=\textwidth]{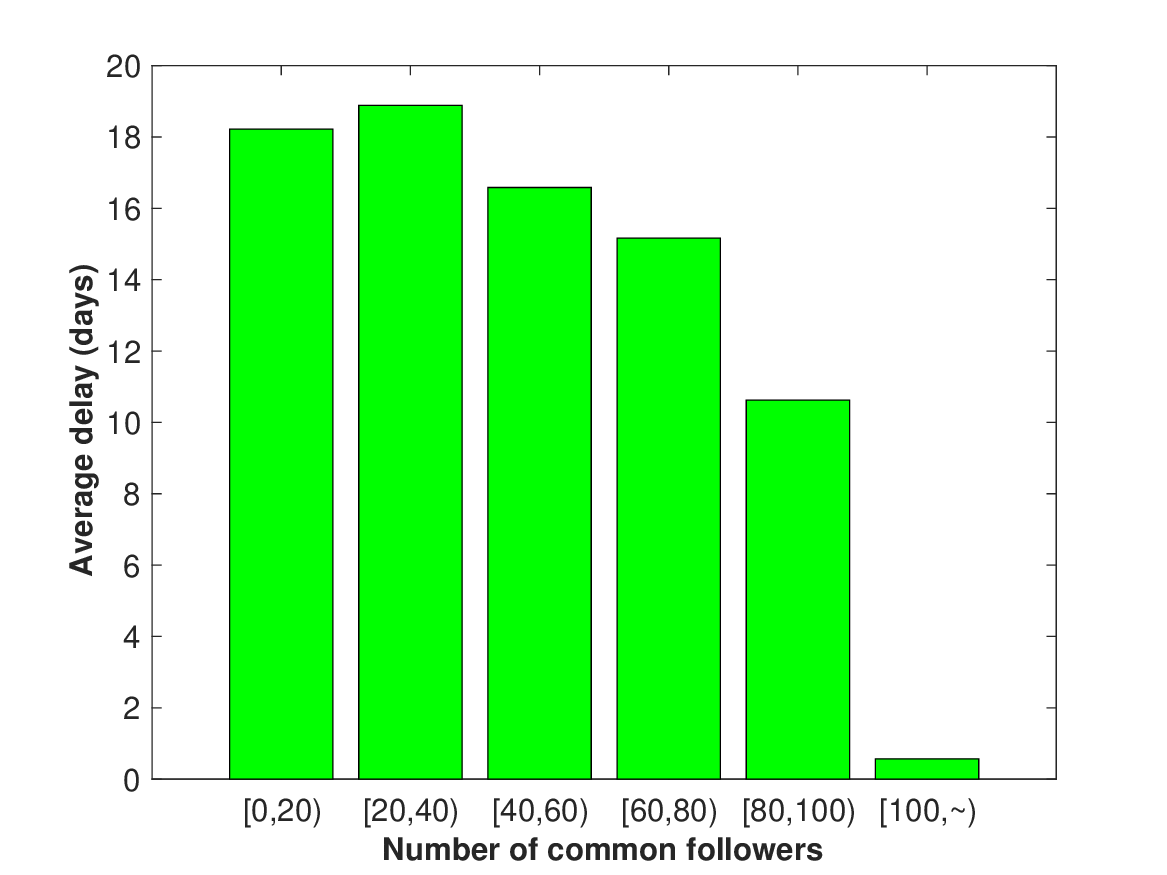}}
\end{minipage}
\caption{Average delay versus the number of common followees and followers.}
\label{fig:co}
\end{figure}

{\it The impact of structural similarity:} Here we engage in some investigations the impacts of some structural similarities such as the number of common followees and the number of common followers for the source and target users at the reciprocity initiate time. In particular, we divide the number of common followees and followers into 6 ranges: $[0,20)$, $[20,40)$, $[40,60)$, $[60,80)$, $[80,100)$ and $[100,+ \infty)$. Their effects on the reciprocal delays are illustrated in Figure~\ref{fig:co}. With the increase of structural similarity between two users (either common followees or common followers), the average delay tends to decrease. For example, when the number of common followers increases to 100, the average delay is reduced to only 1 day compared to more than 18 days with less than $20$ common followers. Social homophily suggests that similar users are likely to create links to each other~\cite{mcpherson2001birds}. Our study extends the social homophily to the temporal dimension for reciprocal relations -- similar users are likely to have short delay in reciprocal relations.

\section{Delay Prediction in Reciprocal Relations}
\begin{table}[!htbp]
\centering
\begin{tabular}{|c|c|}\hline
\multirow{7}{*}{Temporal} & Source user joining time $t_{u}$ \\ \cline{2-2}
    						     & Target user joining time $t_{v}$ \\ \cline{2-2}
    					             & Length of source user joining time $t_{1}-t_{u}$  \\ \cline{2-2}
    						    & Length of target user joining time $t_{1}-t_{v}$\\ \cline{2-2}
						    & Does the reciprocity start on weekends? \\ \cline{2-2}
						    & Avg of previous $k$ delays of user $v$\\ \cline{2-2}
						    & Avg of previous all delays of user $v$\\ \hline\hline
\multirow{6}{*}{Structural }      & Indegree of user $u$ at $t_{1}$\\ \cline{2-2}
    						   &  Indegree of user $v$ at $t_{1}$\\ \cline{2-2}
					            & Outdegree of user $u$ at $t_{1}$ \\ \cline{2-2}
    						    & Outdegree of user $v$ at $t_{1}$  \\ \cline{2-2}
						    & Number of common followees at $t_{1}$\\\cline{2-2}
    							& Number of common followers at $t_{1}$   \\\hline
\end{tabular}
\caption{Summary of features for each reciprocal relation $<u,v,t_{1},?,?-t_{1}>$ where $?$ indicates the following back time.}
\label{table:feature}
\end{table}
It needs to mention that the problem of whether will you follow back or not has been extensively studied in literature~\cite{hopcroft2011will, cheng2011predicting} and these methods have shown to achieve very high performance. However, this is the focus of this study, in this work, we perform further analysis on the problem of when will user follow back on the basis of their work. In other words, if you follow back, we attempt to predict the reciprocal delay.

According to our observations and findings in the previous section, we reveal some interesting patterns that influence the length of the delay in reciprocal relations. Now we attempt to answer the second question whether we can leverage these patterns to predict the delay automatically. In particular, we first extract features that may potentially affect the delay according to our previous analysis. For each reciprocal relation $<u,v,t_{1},?,?-t_{1}>$ (where $?$ is the following back time we want to predict), we extract two types of features that are corresponding to two types of patterns that are summarized in Table~\ref{table:feature}. It should be noted that there might be other factors that may affect the reciprocal delays, but we focus on temporal and structural related factors and leave further investigations as future work. With the feature representation for each reciprocal relation, we consider the delay prediction problem as a regression problem. Given a reciprocal relation $<u,v,t_{1},?,?-t_{1}>$, our goal is to predict the number of days the target user $v$ takes to follow back to the source user $u$. In other words, we would like to train a predictive learning model with the training data where the missing $?$ are already known and then predict the missing follow back time $?$ in each reciprocity relation in the test data. The problem statement of the delay prediction is formally stated as follows: \emph{Given the feature matrix $\mat{X}\in \mathbb{R}^{n\times d}$ where $n$ is the number of training reciprocal relations, and their corresponding delay vector $\mat{y}\in\mathbb{R}^{n}$, we aim to learn to a regression model $f$ that maps $\mat{X}$ to $\mat{y}$}.

\subsection{The Proposed Delay Prediction Framework}
A simple and straightforward way is to build a global model for all reciprocal relations. However, one drawback is that it assumes that reciprocal relations from all users share the exactly same patterns. Though the delay overall tends to follow some common patterns in the previous section, reciprocal behaviors of different individuals could differ. For example, some users may regularly log in and reciprocate back to their followers very quickly while some other users could be not active and spend longer to follow back. Therefore, it is more appealing to build a model to capture both the common patterns and the personalized nature of reciprocal behaviors. Next, we will first introduce the framework to model the common patterns and then extend it to model the personalized nature.

Let $\mat{X}=[\mat{x}_{1},\mat{x}_{2},...,\mat{x}_{n}]^{T}$ ($\mat{x}_{i}\in\mathbb{R}^{d}$, $i=1,...,n$) be the set all $n$ reciprocal relations in the training data. Each reciprocal relation $\mat{x}_{i}$ is also associated with a tuple $<u,v,t_{1},t_{2},t_{2}-t_{1}>$ recording its source user $u$, the target user $v$, the initiating time, the completing time, and the delay, respectively. Assume that the total number of the target users is $K$, $K$ is normally smaller than $n$ since each user may reciprocate back to multiple other users. In particular, we define a matrix $\mat{A}\in\{0,1\}^{n\times n}$ such that $\mat{A}_{ij}=1$ if the target users are the same for two reciprocal relations $\mat{x}_{i}$ and $\mat{x}_{j}$, otherwise $\mat{A}_{ij}=0$. With these definitions, we introduce the framework that captures common patterns. We assume that there is a universal regression parameter $\mat{w}$ between each reciprocal relation $\mat{x}_{i}$ and its corresponding delay $\mat{y}_{i}$. In other words, the delay can be approximated by a ridge regression~\cite{hoerl1970ridge} model:
\begin{equation}
\min_{\mat{w}}\sum_{i=1}^{n}(\mat{x}_{i}^{T}\mat{w}-\mat{y}_{i})^{2}+\alpha\|\mat{w}\|_{2}^{2},
\label{eq:ridge}
\end{equation}
where $\mat{w}\in\mathbb{R}^{d}$ is the regression coefficient, and the term $\alpha\|\mat{w}\|_{2}^{2}$ prevents the overfitting of the model. The model parameter $\mat{w}$ is used to capture the common patterns of all reciprocal relations. Different reciprocal relations could act differently, however, the ridge regression model in Eq.~(\ref{eq:ridge}) cannot fully capture individual reciprocal behaviors of users. To leverage personalized behaviors for accurate delay prediction, we assume that the reciprocal behaviors of each reciprocal relation $\mat{x}_{i}$ is also determined by a local variable $\tilde{\mat{w}}_{i}$ that can seize its personalized behavior. In this way, the delay prediction model can be formulated as:
\begin{equation}
\min_{\mat{w},\tilde{\mat{w}_{i}}}\sum_{i=1}^{n}(\mat{x}_{i}^{T}(\mat{w}+\tilde{\mat{w}}_{i})-\mat{y}_{i})^{2}+\alpha\|\mat{w}\|_{2}^{2}.
\label{eq:personalized}
\end{equation}
In many cases, each user may reciprocate back to a number of other users. In other words, for many reciprocal relations $\mat{x}_{i}$, the target user $v$ are the same. Hence, it is of vital importance to capture the inherent correlations between these reciprocal relations if they share the same user. Particularly, we impose a constraint to force the local variables of two reciprocal relations to be close if they are from the same user. Mathematically, it can be achieved by imposing a network lasso penalization term~\cite{hallac2015network}:
\begin{sequation}
\sum_{i=1}^{n}\sum_{j=1}^{n}\mat{A}_{ij}\|\tilde{\mat{w}}_{i}-\tilde{\mat{w}}_{j}\|_{2}.
\label{eq:networklasso}
\end{sequation}

The advantages of the introduction of the network lasso penalization term are two-fold. First, the $\ell_{2}$-norm penalization term in Eq.~(\ref{eq:networklasso}) not only makes $\tilde{\mat{w}}_{i}$ close to $\tilde{\mat{w}}_{j}$, but incentivizes them exactly to be the same if the corresponding $\mat{A}_{ij}=1$. Second, the $\ell_{2}$-norm penalization term could result in learning more accurate $\tilde{\mat{w}}_{i}$ -- in addition to $\mat{x}_{i}$ as Eq.~(\ref{eq:personalized}), the network lasso term in Eq.~(\ref{eq:networklasso}) also enables us to learning $\tilde{\mat{w}}_{i}$ from related reciprocal relations (e.g., $\mat{x}_{j}$)  that share the same target user with $\mat{x}_{i}$ (i.e., $\mat{A}_{ij} = 1$).

Integrating the formulation in Eq.~\ref{eq:personalized} and Eq.~\ref{eq:networklasso}, we obtain the objective function of the proposed framework for delay prediction in reciprocal relations (DPRR) as follows:
\begin{equation}
\begin{split}
\min_{\mat{w},\tilde{\mat{w}_{i}}}\sum_{i=1}^{n}(\mat{x}_{i}^{T}(\mat{w}+\tilde{\mat{w}}_{i})-\mat{y}_{i})^{2}+\alpha \|\mat{w}\|_{2}^{2}+\beta \sum_{i,j}\mat{A}_{ij}\|\tilde{\mat{w}}_{i}-\tilde{\mat{w}}_{j}\|_{2}
\end{split}
\label{eq:gdp}
\end{equation}
where $\beta$ is to control the consensus of all localized parameters $\tilde{\mat{w}}$. For example, if the parameter $\beta\rightarrow \infty$, all the localized parameters tend to achieve a consensus solution such that $\forall i, j$, we have $\tilde{\mat{w}_{i}}=\tilde{\mat{w}_{j}}$. On the contrary if $\beta=0$, we eliminate the correlations among reciprocal relations and each localized parameter will be learned independently.

By solving the above optimization problem for the proposed framework DPRR , we obtain the universal regression parameter $\mat{w}$ and localized parameters $\tilde{\mat{w}_{i}}$ for each reciprocal relation $\mat{x}_{i}$. Now we discuss how to make a delay prediction for a new reciprocal relation. There are two cases that we need to discuss. The first case is that for the new relation $<u,v,t_{1},?,?-t_{1}>$, we do not find any historical reciprocal behaviors for the target user $v$ in the training data. In this case, we cannot obtain a localized prediction model for the new relation. Hence, we leverage the universal regression parameter $\mat{w}$ to make the delay prediction as $\mat{x}_{k}^{T}\mat{w}$. In the second case, if we find historical reciprocal behaviors for the target node $v$, the localized regression parameter $\tilde{\mat{w}_{k}}$ can be obtained by leveraging all of the localized parameters of all reciprocal relations with the target user $v$. Specifically, we obtain $\tilde{\mat{w}_{k}}$ by solving a Weber problem~\cite{kulin1962efficient}:
\begin{equation}
\tilde{\mat{w}_{k}} = \min_{\mat{b}}\sum_{i=1}^{n}\mat{A}_{ki}{}\|\mat{b}-\tilde{\mat{w}_{i}}\|_{2},
\end{equation}
where $\mat{A}_{ki}=1$ if the target user of $\mat{x}_{i}$ is also $v$. With the solution of $\tilde{\mat{w}_{k}}$, the delay is predicted with the universal parameter $\mat{w}$ and the localized parameter $\tilde{\mat{w}_{k}}$ as $\mat{x}_{k}^{T}(\mat{w}+\tilde{\mat{w}_{k}})$. In the following subsection, we detail the algorithm to optimize the proposed framework DPRR.

\subsection{The Optimization Algorithm}
The objective function of the proposed model in Eq.~(\ref{eq:gdp}) has two sets of parameters -- the universal parameter $\mat{w}$ for all reciprocal relations, and a localized parameter $\tilde{\mat{w}}_{i}$ for each individual reciprocal relation $\mat{x}_{i}$. Motivated by~\cite{hallac2015network}, we use Alternating Direction Method of Multipliers (ADMM)~\cite{boyd2011distributed} to solve it. To make it solvable by ADMM, we design some auxiliary variables, i.e., copies of $\tilde{\mat{w}_{i}}$ to formulate Eq.~(\ref{eq:gdp}) to its equivalent form:
\begin{equation}
\begin{split}
\min_{\mat{a},\mat{w},\mat{z}}&\sum_{i=1}^{n}(\mat{x}_{i}^{T}\mat{a}_{i}-\mat{y}_{i})^{2}+\alpha\|\mat{w}\|_{2}^{2}+\beta \sum_{i,j}\mat{A}_{ij}\|\mat{z}_{ij}-\mat{z}_{ji}\|_{2}\\
& \mbox{s.t.   } \mat{a}_{i} = \mat{z}_{ij}+\mat{w}, \mbox{  } i = 1,...,n,
\end{split}
\end{equation}
where the auxiliary variables $\mat{z}_{ij}$ $(j=1,...,n)$ are copies of $\tilde{\mat{w}_{i}}$. It then can be solved by the following ADMM problem:
\begin{equation}
\begin{split}
&\min_{\mat{a},\mat{w},\mat{z},\mat{u}}L_{\rho}(\mat{a},\mat{w},\mat{z},\mat{u})=\sum_{i=1}^{n}(\mat{x}_{i}^{T}\mat{a}_{i}-\mat{y}_{i})^{2}+\alpha\|\mat{w}\|_{2}^{2}\\
+&\beta\sum_{i,j}\mat{A}_{ij}\|\mat{z}_{ij}-\mat{z}_{ji}\|_{2}-\frac{\rho}{2}\sum_{i,j}\mat{A}_{ij}(\|\mat{u}_{ij}\|_{2}^{2}+\|\mat{u}_{ji}\|_{2}^{2}\\
-&\|\mat{a}_{i}-\mat{z}_{ij}-\mat{w}+\mat{u}_{ij}\|_{2}^{2}-\|\mat{a}_{j}-\mat{z}_{ji}-\mat{w}+\mat{u}_{ji}\|_{2}^{2}),
\end{split}
\end{equation}
where $\rho>0$ is the penalty parameter and $\mat{u}$ is the scaled dual variable. Then the model parameters in each iteration can be updated by following rules iteratively until the objective function converges.

{\it Update $\mat{a}$:} To update $\mat{a}$, we fix all the other variables and remove terms that are irrelevant to $\mat{a}$. Then each $\mat{a}_{i}$ can be updated by solving the following optimization problem:
\begin{equation}
\begin{split}
&\mat{a}_{i}^{k+1} = \argmin_{\mat{a}_{i}} (\mat{x}_{i}^{T}\mat{a}_{i}-\mat{y}_{i})^{2}+\frac{\rho}{2}\sum_{j}\mat{A}_{ij}\\
&(\|\mat{a}_{i}-\mat{z}_{ij}^{k}-\mat{w}^{k}+\mat{u}_{ij}^{k}\|_{2}^{2}+\|\mat{a}_{j}-\mat{z}_{ji}^{k}-\mat{w}^{k}+\mat{u}_{ji}^{k}\|_{2}^{2})
\end{split}
\end{equation}

{\it Update $\mat{w}$:}
Similarly, we fix the other variables except $\mat{w}$, then $\mat{w}$ can be updated by solving the following problem:
\begin{equation}
\begin{split}
\mat{w}^{k+1} = \argmin_{\mat{w}}\frac{\rho}{2}&\sum_{i,j}\mat{A}_{ij}(\|\mat{a}_{i}^{k+1}-\mat{z}_{ij}^{k}-\mat{w}+\mat{u}_{ij}^{k}\|_{2}^{2}\\
+&\|\mat{a}_{j}^{k+1}-\mat{z}_{ji}^{k}-\mat{w}+\mat{u}_{ji}^{k}\|_{2}^{2})+\alpha\|\mat{w}\|_{2}^{2}
\end{split}
\end{equation}

{\it Update $\mat{z}$:}
The parameter $\mat{z}_{ij}$ is a copy of $\tilde{\mat{w}_{i}}$, if $\mat{A}_{ij}=0$ then the corresponding $\mat{z}_{ij}$ is also zero. In other words, we do not need to optimize $\mat{z}_{ij}$. On the other hand if $\mat{A}_{ij}=1$ (also $\mat{A}_{ji}=1$), the corresponding $\mat{z}_{ij}$ and $\mat{z}_{ji}$ are optimized jointly by solving the following objectives:
\begin{equation}
\begin{split}
\mat{z}_{ij}^{k+1},\mat{z}_{ji}^{k+1}=\argmin_{\mat{z}_{ij},\mat{z}_{ji}}\beta&\sum_{i,j}\mat{A}_{ij}\|\mat{z}_{ij}-\mat{z}_{ji}\|_{2}\\
+\frac{\rho}{2}\sum_{i,j}\mat{A}_{ij}&(\|\mat{a}_{i}^{k+1}-\mat{z}_{ij}-\mat{w}^{k+1}+\mat{u}_{ij}^{k}\|_{2}^{2}\\
+&\|\mat{a}_{j}^{k+1}-\mat{z}_{ji}-\mat{w}^{k+1}+\mat{u}_{ji}^{k}\|_{2}^{2}).
\end{split}
\end{equation}

{\it Update $\mat{u}$:}
At last, in the dual ascent step, we update the scaled dual variable $\mat{u}$ as:
\begin{equation}
\mat{u}_{ij}^{k+1}=\mat{u}_{ij}^{k}+(\mat{a}_{i}^{k+1}-\mat{w}^{k+1}-\mat{z}_{ij}^{k+1}).
\end{equation}

\section{Experiments}
In this section, we present experiments to show the performance of the proposed DPRR framework for delay prediction in reciprocal relations. We begin by introducing the experimental settings, then presenting the detailed results and finally investigating the parameter sensitivity.

\subsection{Experimental Settings}
It is worth noting that the delay prediction problem is different from the reciprocal prediction problem. Current state-of-the-art algorithms already achieve up to 90\% accuracy in predicting if a user will follow back when another user initiates the following relation first. Considering its high accuracy, we study the problem of ``when user will follow back" in a given reciprocal relation. To be more specific, in the training phase, we use the reciprocal relations whose delays have been already known to train the DPRR model. Also, in the test data, we already know that the user will follow back to another, and focus on predicting the reciprocal delay. Each time we sample 100,000 reciprocal relations for training and then sample another $100,000*x\%$ reciprocal relations for testing, where the value of $x$ is varied as $\{50, 70, 90\}$. The whole process is repeated 10 times, and we report the average RMSE and MAE of these 10 trials as the final results. The parameters for all the baseline methods and the proposed model are determined by cross-validation. We compare the proposed framework DPRR with the following baseline methods:
\begin{itemize}
\item P$1$: the delay is predicted using that of the immediately previous one;
\item P$k$: the delay is predicted using the average delay of previous $k$ delays.  Note that in the experiment, we try various $k$ and report the best performance;
\item RG: the delay is predicted with ridge regression model on features extracted via our analysis in Table~\ref{table:feature}. It is a variant of the proposed framework without the model component to capture the personalized nature of reciprocal relations;
\item LS: the delay is predicted with lasso model on features in Table~\ref{table:feature}.
\item PD: It is a variant of the proposed DPRR framework without the model component to capture common patterns. It can be achieved by removing the common parameter $\mat{w}$ from the proposed framework DPRR.
\end{itemize}

\subsection{Performance Comparison of Delay Prediction in Reciprocal Relations}
\begin{figure}[!t]
\centering
\begin{minipage}{0.48\textwidth}
\centering
\subfigure[MAE\label{fig:MAE}]
{\includegraphics[width=\textwidth]{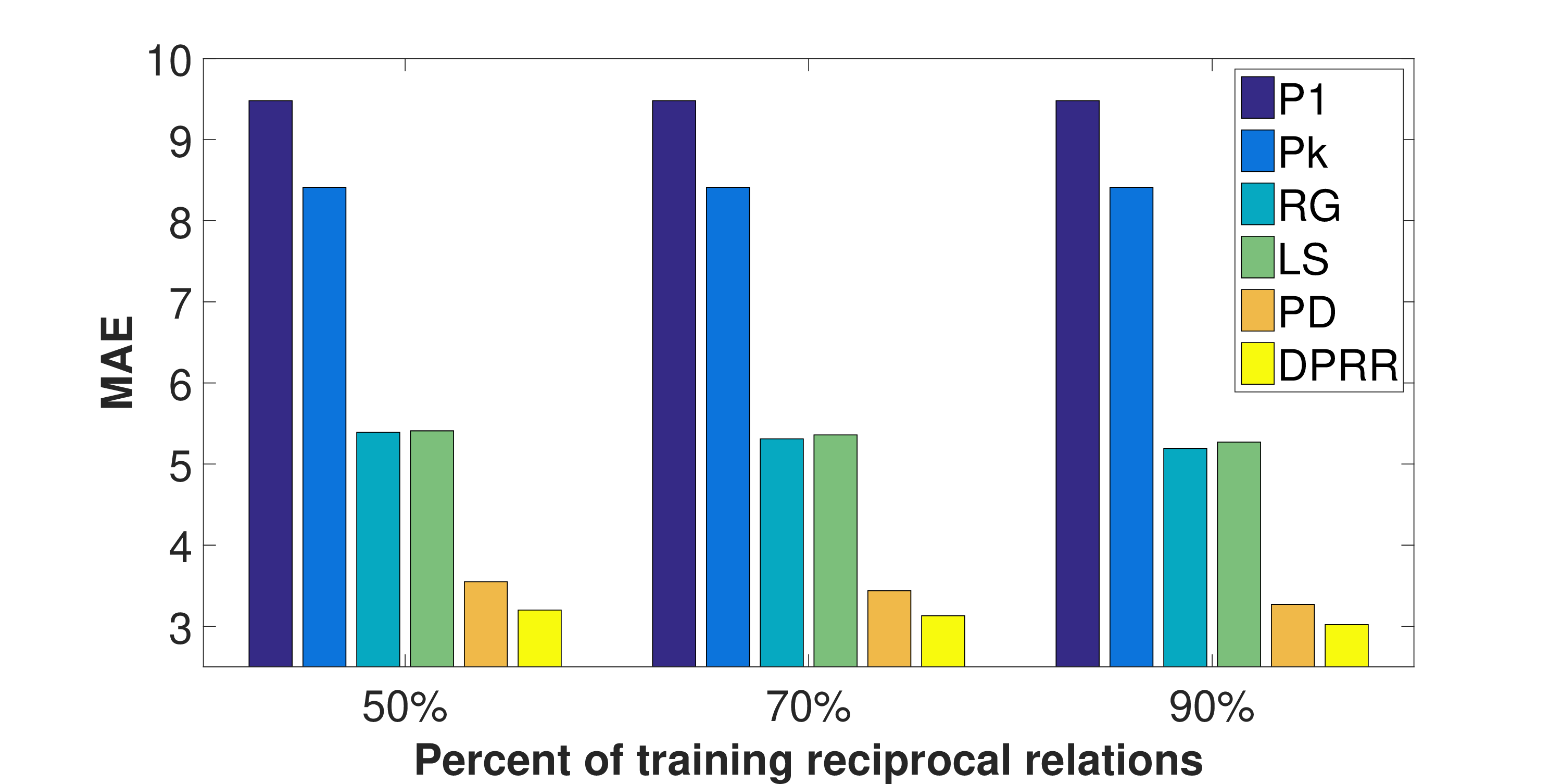}}
\end{minipage}
\begin{minipage}{0.48\textwidth}
\centering
\subfigure[RMSE\label{fig:RMSE}]
{\includegraphics[width=\textwidth]{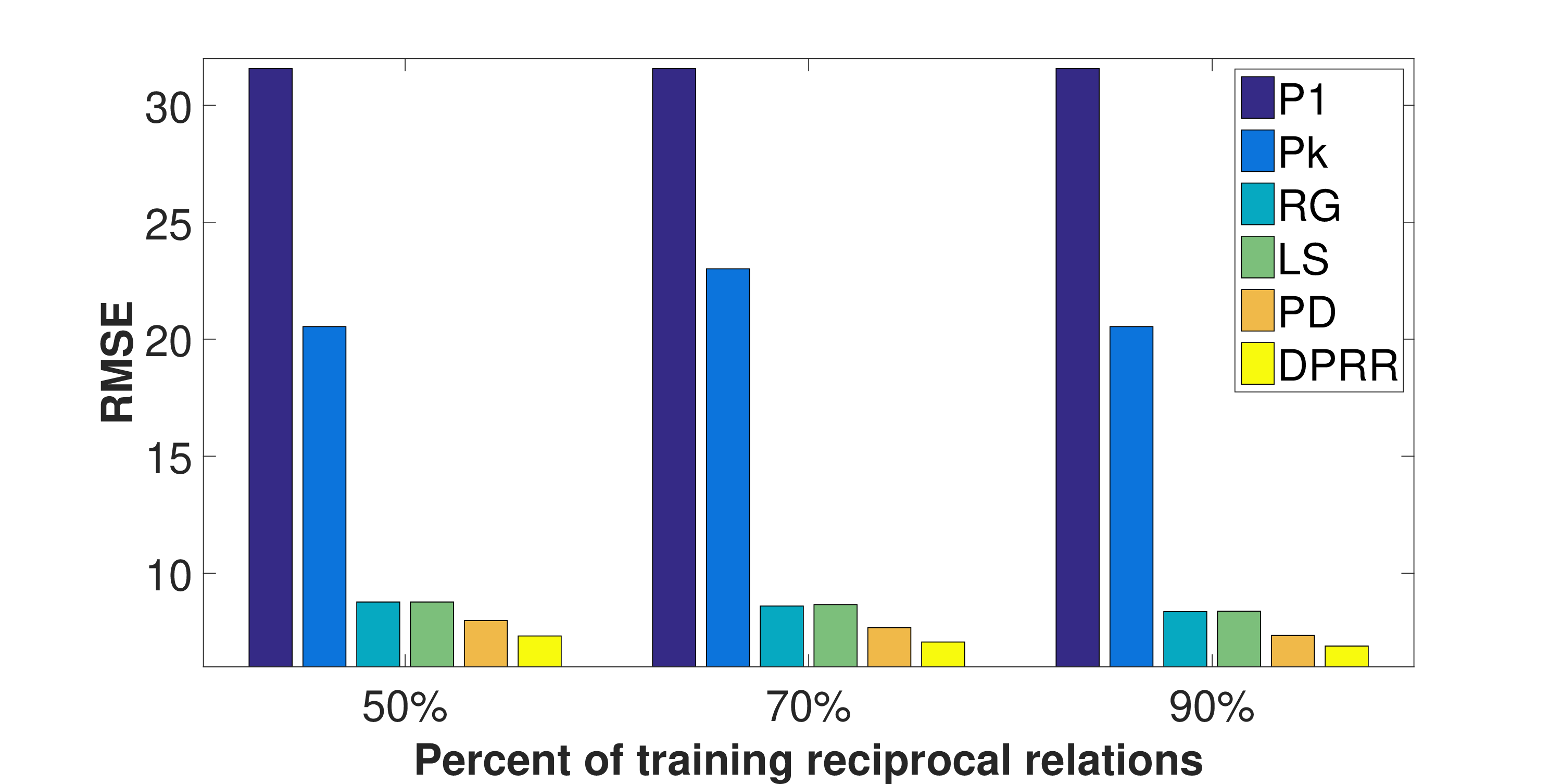}}
\end{minipage}
\caption{Performance of different methods in delay prediction in reciprocal relations.}
\label{fig:performance}
\end{figure}

\begin{table}[!htbp]
\centering
\begin{tabular}{c |c c c} \hline
& &MAE &   \\ \hline
  Test/Training Ratio& 50\% 		& 70\% 		& 90\% 		    \\ \hline \hline
RG & 40.63\%	& 41.05\%	& 41.81\%	 \\ \hline
LS & 40.85\%	& 41.60\%	& 42.69\%	 \\ \hline
PD & 9.86\%		& 9.01\% 	& 7.65\%	\\ \hline \hline
& &RMSE &  \\ \hline
Test/Training Ratio & 50\% 		& 70\% 		& 90\% 		    \\ \hline \hline
RG & 16.53\%	&17.91\%	& 17.58\% \\ \hline
LS & 16.53\%	& 18.48\%	& 17.78\% \\ \hline
PD &  8.27\% 	& 8.07\%	& 6.13\%\\ \hline
\end{tabular}
\caption{The relative performance improvement of the proposed framework compared to baselines.}
\vspace{-0.3in}
\label{table:improvements}
\end{table}

The delay prediction results w.r.t. different sizes of test data are shown in Figure~\ref{fig:performance}. The smaller the values of MAE or RMSE are, the better the performance of the prediction algorithm is. We make the following observations:
\begin{itemize}
\item P$k$ alway outperforms P$1$ that is consistent with our previous observation -- the average of previous $k$ delay is more accurate in predicting the current delay than the immediately previous one;
\item The regression models with features extracted based on the delay patterns achieve much better performance than P$k$. This observation supports the importance to understand the delay patterns;
\item The proposed model DPRR outperforms all baseline methods. We also perform t-test between DPRR and baseline methods, and the test results show that DPRR is significantly better, with a 0.05 significance level.  The relative performance improvement of DPRR compared to regression-based baselines is demonstrated in Table~\ref{table:improvements}. The major reason is that DPRR captures the common patterns via the shared parameter ${\bf w}$ and the personalized nature of each reciprocal relation via a localized parameter $\tilde{\mat{w_{i}}}$  into a coherent model;
\item DPRR obtains much better performance than its two variants as shown in Table~\ref{table:improvements}.  The improvement of DPRR over RG supports the importance of modeling the personalized nature; while the improvement of DPRR over PD suggests the importance of modeling common patterns.
\end{itemize}
With these observations, we can conclude that the proposed DPRR framework can accurately predict delay in reciprocal relations by exploiting both personalized characteristics and common patterns of reciprocal relations.

\subsection{Parameter Selection}
\begin{figure}[!t]
\centering
\begin{minipage}{0.235\textwidth}
\centering
\subfigure[MAE\label{fig:MAE}]
{\includegraphics[width=\textwidth]{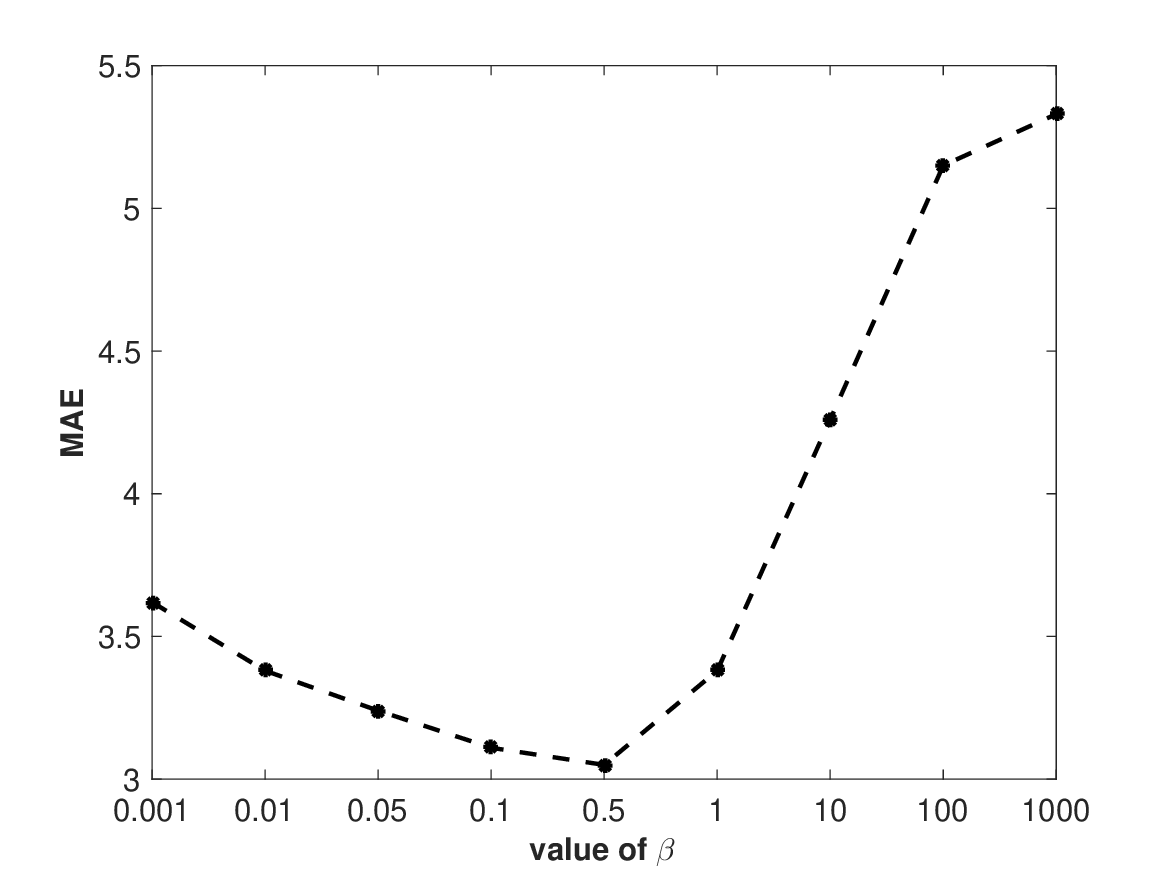}}
\end{minipage}
\begin{minipage}{0.235\textwidth}
\centering
\subfigure[RMSE\label{fig:RMSE}]
{\includegraphics[width=\textwidth]{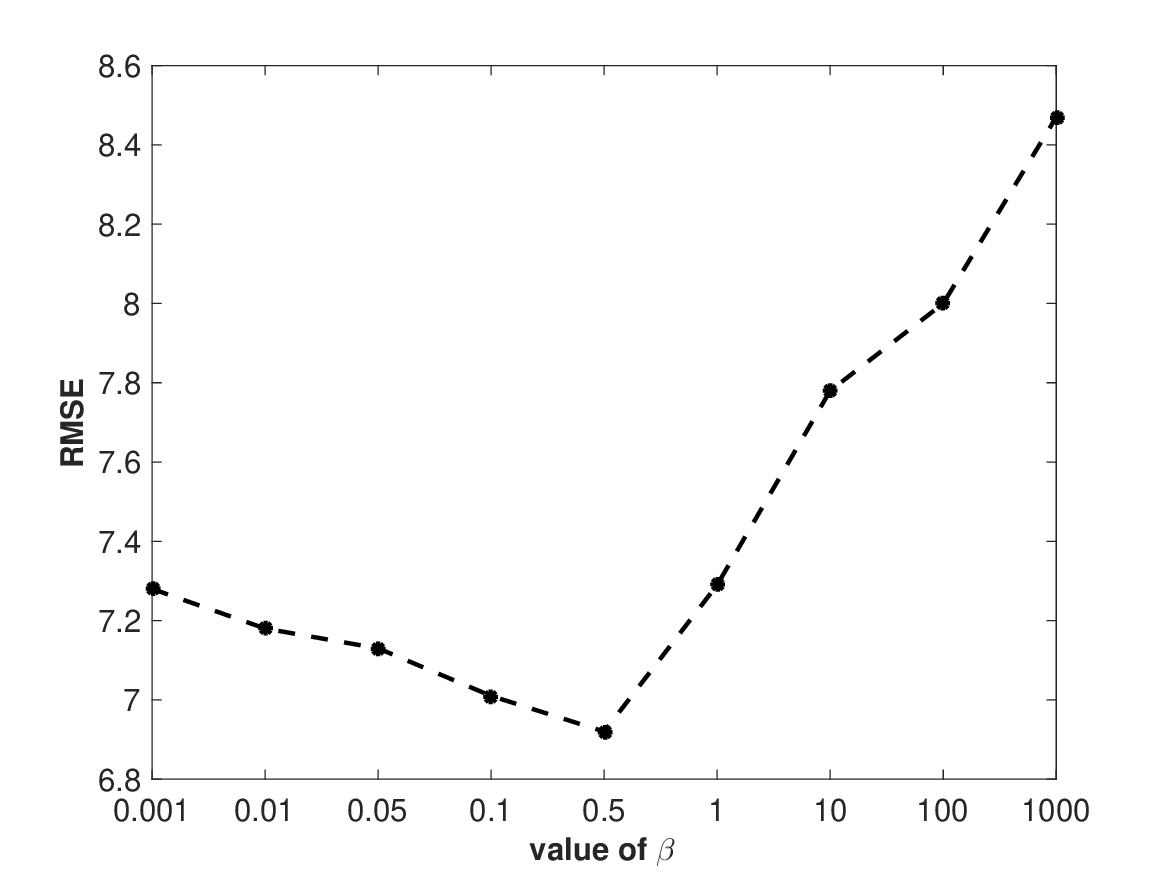}}
\end{minipage}
\caption{Performance variation w.r.t. $\beta$ in DPRR.}
\vspace{-0.2in}
\label{fig:beta}
\end{figure}

The proposed framework has one important pre-defined parameter $\beta$ that controls the contribution of the network lasso term. In this subsection, we investigate its impact by investigating how the performance varies with the changes of the value of $\beta$.  We try $\beta \in \{0.001,0.01,0.05,0.1,0.5,1,10$, $100,1000\}$ and the performance of DPRR w.r.t. $\beta$ is shown in the Figure~\ref{fig:beta}. In the figure, we make the following observations -- (1) by gradually increasing the value of $\beta$, MAE and RMSE tend to decrease and then increase, and the best performance is achieved when $\beta$ is around 0.5; (2) When $\beta$ is very small, each reciprocal relation inclines to have its own regularized parameter; on the other hand, when $\beta$ is very large, all reciprocal relations tend to share a common regularized parameter. MAE and RMSE are relatively higher in these two cases, showing the necessity of leveraging personalized and global patterns for delay prediction.

\section{Related Work}
In this section, we briefly review related work from two aspects: (1) reciprocity behaviors analysis and prediction; (2) link prediction with time information.
\subsection{Reciprocity Analysis and Prediction}
The study of link reciprocity has received increasing attention recently. Garlaschelli and Loffredo~\cite{garlaschelli2004patterns} propose a new measure of reciprocity to unveil reciprocity behaviors in a variety of real-world networks. Squartini et al.~\cite{squartini2013reciprocity} extend the current reciprocity measure to weighted networks. Also, they show that reciprocity links play a central role in the dynamic process, network growth, and community evolution of a network. Leman et al.~\cite{akoglu2012quantifying} quantify the degree of reciprocity in weighted networks and study its correlation with user pair topological features. Nguyen et al.~\cite{nguyen2010you} investigate to what extent the reciprocity relations exist in a trust network. They study four types of user reciprocity behaviors and propose a learning model to perform the reciprocity trust prediction. Hopcroft et al. leverage~\cite{hopcroft2011will} social theories to predict reciprocal relationships in a microblogging platform Twitter. The proposed factor graph model can accurately predict more than 90\% of reciprocal relationships. Cheng et al.~\cite{cheng2011predicting} formulate the reciprocity relationship prediction problem as a supervised learning task and identify a set of informative features beneficial for the reciprocity prediction problem. Methods in~\cite{hopcroft2011will, cheng2011predicting} need negative unreciprocated training samples that may turn into reciprocal relations in the future, thus the generalization of their methods is limited. Instead, in~\cite{gong2013reciprocity}, the reciprocity prediction problem is modeled as an outlier detection problem. Feng et al.~\cite{feng2014time} show that the time delay of the reciprocity relations also has a significant impact on the reciprocity formation. Therefore, they propose a time-aware reciprocity relation prediction model by employing the reciprocity time delay. All the above mentioned methods are distinct from the goal of this work as they either focus on (1) studying or measuring reciprocity relations; or (2) performing the reciprocity relation prediction task. In other words, they fail to assess the delay in reciprocal relations explicitly and cannot answer the question ``when" will users follow you back.

\subsection{Link Prediction with Time Information}
Another line of work that are remotely related the problem we study is link prediction with time information. Link prediction problem targets to predict the future potential links mainly based on the snapshot of current network structure~\cite{liben2007link}. Despite its importance, the prediction of when a link will appear in the new future remains an under-explored task. Sun et al.~\cite{sun2012will} make one of the first attempt to predict when will the links be created in heterogeneous information networks with a meta-path based approach. Li et al.~\cite{li2016predicting} propose a time difference labeled based approach to capture the correlation between structural edges and time information in predicting the links and their building time. There are also a number of other methods which do not explicitly predict the building time of links but exploit the temporal information to improve the performance of the link prediction. For example, Sakar et al.~\cite{sarkar2012nonparametric} propose a nonparametric prediction algorithm for a sequence of network structure snapshots over time. In~\cite{tylenda2009towards} and~\cite{huang2009time}, they show that incorporation of historical information of user interactions can significantly improve the link prediction accuracy. In~\cite{kolar2010estimating}, the time-varying link prediction problem is formulated by a temporally smoothed $\ell_{1}$ logistic regression problem. Li et al.~\cite{li2016predicting} propose a time difference labeled based approach to capture the correlation between structural edges and time information in predicting the links and their building time. In addition, this work is also remotely related to the link prediction problem on dynamic networks~\cite{dunlavy2011temporal,zhu2016scalable,yu2017link,li2018streaming}. 

\section{Conclusions and Future Work}
In directed social networks, reciprocity represents bidirectional relations among two users, showing the willings of one user to return a favor impacted by the other. Reciprocal relations create closer social ties, and its time delay is essential in understanding the formation of user interactions, network growth and dynamics. This paper presents the first study of delay in reciprocal relations with a large-scale dynamic network from Tumblr. We study some potential factors that may influence the length of the delay and find that the delay shows some interesting temporal and structural patterns.  Meanwhile, we investigate how to leverage these observations into a principled model to automatically predict delay in reciprocal relations. Methodologically, we propose a DPRR model to capture some common and personalized patterns of reciprocal relations. Empirical experiments on a large-scale dataset validate the effectiveness of the proposed model.

Future research can be focused on two directions. First, the understanding and predicting of delay in reciprocal relations may benefit various of applications such as friend recommendation, viral marketing, etc. Thus we plan to investigate how to deploy our findings and the prediction model on these applications. Second, this work makes use of the temporal and structural information for the delay prediction task on dynamic networks. In social media, user generated and user interaction contents are prevalent and continuously evolve over time~\cite{li2016toward,li2017attributed}, we would like to study how to incorporate other rich sources of information in the prediction model to improve its performance.

\section*{Acknowledgements}
This material is based upon work supported by, or in part by, the National Science Foundation (NSF) grant 1614576, and the Office of Naval Research (ONR) grant N00014-16-1-2257.

\balance


\begin{thebibliography}{44}


\ifx \showCODEN    \undefined \def \showCODEN     #1{\unskip}     \fi
\ifx \showDOI      \undefined \def \showDOI       #1{#1}\fi
\ifx \showISBNx    \undefined \def \showISBNx     #1{\unskip}     \fi
\ifx \showISBNxiii \undefined \def \showISBNxiii  #1{\unskip}     \fi
\ifx \showISSN     \undefined \def \showISSN      #1{\unskip}     \fi
\ifx \showLCCN     \undefined \def \showLCCN      #1{\unskip}     \fi
\ifx \shownote     \undefined \def \shownote      #1{#1}          \fi
\ifx \showarticletitle \undefined \def \showarticletitle #1{#1}   \fi
\ifx \showURL      \undefined \def \showURL       {\relax}        \fi
\providecommand\bibfield[2]{#2}
\providecommand\bibinfo[2]{#2}
\providecommand\natexlab[1]{#1}
\providecommand\showeprint[2][]{arXiv:#2}

\bibitem[\protect\citeauthoryear{Aggarwal and Subbian}{Aggarwal and
  Subbian}{2014}]%
        {aggarwal2014evolutionary}
\bibfield{author}{\bibinfo{person}{Charu Aggarwal} {and}
  \bibinfo{person}{Karthik Subbian}.} \bibinfo{year}{2014}\natexlab{}.
\newblock \showarticletitle{Evolutionary network analysis: A survey}.
\newblock \bibinfo{journal}{\emph{Comput. Surveys}} \bibinfo{volume}{47},
  \bibinfo{number}{1} (\bibinfo{year}{2014}), \bibinfo{pages}{10}.
\newblock


\bibitem[\protect\citeauthoryear{Aggarwal and Subbian}{Aggarwal and
  Subbian}{2012}]%
        {aggarwal2012event}
\bibfield{author}{\bibinfo{person}{Charu~C Aggarwal} {and}
  \bibinfo{person}{Karthik Subbian}.} \bibinfo{year}{2012}\natexlab{}.
\newblock \showarticletitle{Event Detection in Social Streams.}. In
  \bibinfo{booktitle}{\emph{SDM}}. SIAM, \bibinfo{pages}{624--635}.
\newblock


\bibitem[\protect\citeauthoryear{Akoglu, de~Melo, and Faloutsos}{Akoglu
  et~al\mbox{.}}{2012}]%
        {akoglu2012quantifying}
\bibfield{author}{\bibinfo{person}{Leman Akoglu}, \bibinfo{person}{Pedro OS~Vaz
  de Melo}, {and} \bibinfo{person}{Christos Faloutsos}.}
  \bibinfo{year}{2012}\natexlab{}.
\newblock \showarticletitle{Quantifying reciprocity in large weighted
  communication networks}. In \bibinfo{booktitle}{\emph{PAKDD}}. Springer,
  \bibinfo{pages}{85--96}.
\newblock


\bibitem[\protect\citeauthoryear{Boyd, Parikh, Chu, Peleato, and Eckstein}{Boyd
  et~al\mbox{.}}{2011}]%
        {boyd2011distributed}
\bibfield{author}{\bibinfo{person}{Stephen Boyd}, \bibinfo{person}{Neal
  Parikh}, \bibinfo{person}{Eric Chu}, \bibinfo{person}{Borja Peleato}, {and}
  \bibinfo{person}{Jonathan Eckstein}.} \bibinfo{year}{2011}\natexlab{}.
\newblock \showarticletitle{Distributed optimization and statistical learning
  via the alternating direction method of multipliers}.
\newblock \bibinfo{journal}{\emph{Foundations and Trends in Machine Learning}}
  \bibinfo{volume}{3}, \bibinfo{number}{1} (\bibinfo{year}{2011}),
  \bibinfo{pages}{1--122}.
\newblock


\bibitem[\protect\citeauthoryear{Chang and Sun}{Chang and Sun}{2011}]%
        {chang2011location}
\bibfield{author}{\bibinfo{person}{Jonathan Chang} {and} \bibinfo{person}{Eric
  Sun}.} \bibinfo{year}{2011}\natexlab{}.
\newblock \showarticletitle{Location 3: How users share and respond to
  location-based data on social networking sites}. In
  \bibinfo{booktitle}{\emph{ICWSM}}. \bibinfo{pages}{74--80}.
\newblock


\bibitem[\protect\citeauthoryear{Chang, Tang, Inagaki, and Liu}{Chang
  et~al\mbox{.}}{2014}]%
        {chang2014tumblr}
\bibfield{author}{\bibinfo{person}{Yi Chang}, \bibinfo{person}{Lei Tang},
  \bibinfo{person}{Yoshiyuki Inagaki}, {and} \bibinfo{person}{Yan Liu}.}
  \bibinfo{year}{2014}\natexlab{}.
\newblock \showarticletitle{What is tumblr: A statistical overview and
  comparison}.
\newblock \bibinfo{journal}{\emph{ACM SIGKDD Explorations Newsletter}}
  \bibinfo{volume}{16}, \bibinfo{number}{1} (\bibinfo{year}{2014}),
  \bibinfo{pages}{21--29}.
\newblock


\bibitem[\protect\citeauthoryear{Cheng, Adamic, Dow, Kleinberg, and
  Leskovec}{Cheng et~al\mbox{.}}{2014}]%
        {cheng2014can}
\bibfield{author}{\bibinfo{person}{Justin Cheng}, \bibinfo{person}{Lada
  Adamic}, \bibinfo{person}{P~Alex Dow}, \bibinfo{person}{Jon~Michael
  Kleinberg}, {and} \bibinfo{person}{Jure Leskovec}.}
  \bibinfo{year}{2014}\natexlab{}.
\newblock \showarticletitle{Can cascades be predicted?}. In
  \bibinfo{booktitle}{\emph{WWW}}. ACM, \bibinfo{pages}{925--936}.
\newblock


\bibitem[\protect\citeauthoryear{Cheng, Romero, Meeder, and Kleinberg}{Cheng
  et~al\mbox{.}}{2011}]%
        {cheng2011predicting}
\bibfield{author}{\bibinfo{person}{Justin Cheng}, \bibinfo{person}{Daniel~M
  Romero}, \bibinfo{person}{Brendan Meeder}, {and} \bibinfo{person}{Jon
  Kleinberg}.} \bibinfo{year}{2011}\natexlab{}.
\newblock \showarticletitle{Predicting reciprocity in social networks}. In
  \bibinfo{booktitle}{\emph{SocialCom}}. IEEE, \bibinfo{pages}{49--56}.
\newblock


\bibitem[\protect\citeauthoryear{Dunlavy, Kolda, and Acar}{Dunlavy
  et~al\mbox{.}}{2011}]%
        {dunlavy2011temporal}
\bibfield{author}{\bibinfo{person}{Daniel~M Dunlavy}, \bibinfo{person}{Tamara~G
  Kolda}, {and} \bibinfo{person}{Evrim Acar}.} \bibinfo{year}{2011}\natexlab{}.
\newblock \showarticletitle{Temporal link prediction using matrix and tensor
  factorizations}.
\newblock \bibinfo{journal}{\emph{ACM Transactions on Knowledge Discovery from
  Data}} \bibinfo{volume}{5}, \bibinfo{number}{2} (\bibinfo{year}{2011}),
  \bibinfo{pages}{10}.
\newblock


\bibitem[\protect\citeauthoryear{Feng, Zhao, Fang, and Xu}{Feng
  et~al\mbox{.}}{2014}]%
        {feng2014time}
\bibfield{author}{\bibinfo{person}{Xu Feng}, \bibinfo{person}{Jichang Zhao},
  \bibinfo{person}{Zhiwen Fang}, {and} \bibinfo{person}{Ke Xu}.}
  \bibinfo{year}{2014}\natexlab{}.
\newblock \showarticletitle{Time-aware reciprocity prediction in trust
  network}. In \bibinfo{booktitle}{\emph{ASONAM}}. IEEE,
  \bibinfo{pages}{234--237}.
\newblock


\bibitem[\protect\citeauthoryear{Gao, Tang, and Liu}{Gao et~al\mbox{.}}{2012}]%
        {gao2012exploring}
\bibfield{author}{\bibinfo{person}{Huiji Gao}, \bibinfo{person}{Jiliang Tang},
  {and} \bibinfo{person}{Huan Liu}.} \bibinfo{year}{2012}\natexlab{}.
\newblock \showarticletitle{Exploring Social-Historical Ties on Location-Based
  Social Networks.}. In \bibinfo{booktitle}{\emph{ICWSM}}.
  \bibinfo{pages}{114--121}.
\newblock


\bibitem[\protect\citeauthoryear{Garlaschelli and Loffredo}{Garlaschelli and
  Loffredo}{2004}]%
        {garlaschelli2004patterns}
\bibfield{author}{\bibinfo{person}{Diego Garlaschelli} {and}
  \bibinfo{person}{Maria~I Loffredo}.} \bibinfo{year}{2004}\natexlab{}.
\newblock \showarticletitle{Patterns of link reciprocity in directed networks}.
\newblock \bibinfo{journal}{\emph{Physical Review Letters}}
  \bibinfo{volume}{93}, \bibinfo{number}{26} (\bibinfo{year}{2004}),
  \bibinfo{pages}{268701}.
\newblock


\bibitem[\protect\citeauthoryear{Gong, Xu, and Song}{Gong
  et~al\mbox{.}}{2013}]%
        {gong2013reciprocity}
\bibfield{author}{\bibinfo{person}{N~Zhenqiang Gong}, \bibinfo{person}{Wenchang
  Xu}, {and} \bibinfo{person}{Dawn Song}.} \bibinfo{year}{2013}\natexlab{}.
\newblock \showarticletitle{Reciprocity in social networks: Measurements,
  predictions, and implications}.
\newblock \bibinfo{journal}{\emph{ArXiv e-prints}} (\bibinfo{year}{2013}).
\newblock


\bibitem[\protect\citeauthoryear{Hallac, Leskovec, and Boyd}{Hallac
  et~al\mbox{.}}{2015}]%
        {hallac2015network}
\bibfield{author}{\bibinfo{person}{David Hallac}, \bibinfo{person}{Jure
  Leskovec}, {and} \bibinfo{person}{Stephen Boyd}.}
  \bibinfo{year}{2015}\natexlab{}.
\newblock \showarticletitle{Network lasso: Clustering and optimization in large
  graphs}. In \bibinfo{booktitle}{\emph{SIGKDD}}. ACM,
  \bibinfo{pages}{387--396}.
\newblock


\bibitem[\protect\citeauthoryear{Hoerl and Kennard}{Hoerl and Kennard}{1970}]%
        {hoerl1970ridge}
\bibfield{author}{\bibinfo{person}{Arthur~E Hoerl} {and}
  \bibinfo{person}{Robert~W Kennard}.} \bibinfo{year}{1970}\natexlab{}.
\newblock \showarticletitle{Ridge regression: Biased estimation for
  nonorthogonal problems}.
\newblock \bibinfo{journal}{\emph{Technometrics}} \bibinfo{volume}{12},
  \bibinfo{number}{1} (\bibinfo{year}{1970}), \bibinfo{pages}{55--67}.
\newblock


\bibitem[\protect\citeauthoryear{Hopcroft, Lou, and Tang}{Hopcroft
  et~al\mbox{.}}{2011}]%
        {hopcroft2011will}
\bibfield{author}{\bibinfo{person}{John Hopcroft}, \bibinfo{person}{Tiancheng
  Lou}, {and} \bibinfo{person}{Jie Tang}.} \bibinfo{year}{2011}\natexlab{}.
\newblock \showarticletitle{Who will follow you back?: reciprocal relationship
  prediction}. In \bibinfo{booktitle}{\emph{CIKM}}. ACM,
  \bibinfo{pages}{1137--1146}.
\newblock


\bibitem[\protect\citeauthoryear{Horton and Richard~Wohl}{Horton and
  Richard~Wohl}{1956}]%
        {horton1956mass}
\bibfield{author}{\bibinfo{person}{Donald Horton} {and} \bibinfo{person}{R
  Richard~Wohl}.} \bibinfo{year}{1956}\natexlab{}.
\newblock \showarticletitle{Mass communication and para-social interaction:
  Observations on intimacy at a distance}.
\newblock \bibinfo{journal}{\emph{Psychiatry}} \bibinfo{volume}{19},
  \bibinfo{number}{3} (\bibinfo{year}{1956}), \bibinfo{pages}{215--229}.
\newblock


\bibitem[\protect\citeauthoryear{Huang and Lin}{Huang and Lin}{2009}]%
        {huang2009time}
\bibfield{author}{\bibinfo{person}{Zan Huang} {and} \bibinfo{person}{Dennis~KJ
  Lin}.} \bibinfo{year}{2009}\natexlab{}.
\newblock \showarticletitle{The time-series link prediction problem with
  applications in communication surveillance}.
\newblock \bibinfo{journal}{\emph{INFORMS Journal on Computing}}
  \bibinfo{volume}{21}, \bibinfo{number}{2} (\bibinfo{year}{2009}),
  \bibinfo{pages}{286--303}.
\newblock


\bibitem[\protect\citeauthoryear{Jun and Sethi}{Jun and Sethi}{2009}]%
        {jun2009reciprocity}
\bibfield{author}{\bibinfo{person}{Tackseung Jun} {and} \bibinfo{person}{Rajiv
  Sethi}.} \bibinfo{year}{2009}\natexlab{}.
\newblock \showarticletitle{Reciprocity in evolving social networks}.
\newblock \bibinfo{journal}{\emph{Journal of Evolutionary Economics}}
  \bibinfo{volume}{19}, \bibinfo{number}{3} (\bibinfo{year}{2009}),
  \bibinfo{pages}{379--396}.
\newblock


\bibitem[\protect\citeauthoryear{Kolar, Song, Ahmed, and Xing}{Kolar
  et~al\mbox{.}}{2010}]%
        {kolar2010estimating}
\bibfield{author}{\bibinfo{person}{Mladen Kolar}, \bibinfo{person}{Le Song},
  \bibinfo{person}{Amr Ahmed}, {and} \bibinfo{person}{Eric~P Xing}.}
  \bibinfo{year}{2010}\natexlab{}.
\newblock \showarticletitle{Estimating time-varying networks}.
\newblock \bibinfo{journal}{\emph{The Annals of Applied Statistics}}
  (\bibinfo{year}{2010}), \bibinfo{pages}{94--123}.
\newblock


\bibitem[\protect\citeauthoryear{Kulin and Kuenne}{Kulin and Kuenne}{1962}]%
        {kulin1962efficient}
\bibfield{author}{\bibinfo{person}{Harold~W Kulin} {and}
  \bibinfo{person}{Robert~E Kuenne}.} \bibinfo{year}{1962}\natexlab{}.
\newblock \showarticletitle{An efficient algorithm for the numerical solution
  of the generalized Weber problem in spatial economics}.
\newblock \bibinfo{journal}{\emph{Journal of Regional Science}}
  \bibinfo{volume}{4}, \bibinfo{number}{2} (\bibinfo{year}{1962}),
  \bibinfo{pages}{21--33}.
\newblock


\bibitem[\protect\citeauthoryear{Leskovec, Adamic, and Huberman}{Leskovec
  et~al\mbox{.}}{2007}]%
        {leskovec2007dynamics}
\bibfield{author}{\bibinfo{person}{Jure Leskovec}, \bibinfo{person}{Lada~A
  Adamic}, {and} \bibinfo{person}{Bernardo~A Huberman}.}
  \bibinfo{year}{2007}\natexlab{}.
\newblock \showarticletitle{The dynamics of viral marketing}.
\newblock \bibinfo{journal}{\emph{ACM Transactions on the Web}}
  \bibinfo{volume}{1}, \bibinfo{number}{1} (\bibinfo{year}{2007}),
  \bibinfo{pages}{5}.
\newblock


\bibitem[\protect\citeauthoryear{Leskovec, Kleinberg, and Faloutsos}{Leskovec
  et~al\mbox{.}}{2005}]%
        {leskovec2005graphs}
\bibfield{author}{\bibinfo{person}{Jure Leskovec}, \bibinfo{person}{Jon
  Kleinberg}, {and} \bibinfo{person}{Christos Faloutsos}.}
  \bibinfo{year}{2005}\natexlab{}.
\newblock \showarticletitle{Graphs over time: densification laws, shrinking
  diameters and possible explanations}. In \bibinfo{booktitle}{\emph{SIGKDD}}.
  ACM, \bibinfo{pages}{177--187}.
\newblock


\bibitem[\protect\citeauthoryear{Li, Cheng, Wu, and Liu}{Li
  et~al\mbox{.}}{2018}]%
        {li2018streaming}
\bibfield{author}{\bibinfo{person}{Jundong Li}, \bibinfo{person}{Kewei Cheng},
  \bibinfo{person}{Liang Wu}, {and} \bibinfo{person}{Huan Liu}.}
  \bibinfo{year}{2018}\natexlab{}.
\newblock \showarticletitle{Streaming Link Prediction on Dynamic Attributed
  Networks}. In \bibinfo{booktitle}{\emph{WSDM}}. ACM,
  \bibinfo{pages}{369--377}.
\newblock


\bibitem[\protect\citeauthoryear{Li, Dani, Hu, Tang, Chang, and Liu}{Li
  et~al\mbox{.}}{2017}]%
        {li2017attributed}
\bibfield{author}{\bibinfo{person}{Jundong Li}, \bibinfo{person}{Harsh Dani},
  \bibinfo{person}{Xia Hu}, \bibinfo{person}{Jiliang Tang}, \bibinfo{person}{Yi
  Chang}, {and} \bibinfo{person}{Huan Liu}.} \bibinfo{year}{2017}\natexlab{}.
\newblock \showarticletitle{Attributed network embedding for learning in a
  dynamic environment}. In \bibinfo{booktitle}{\emph{CIKM}}. ACM,
  \bibinfo{pages}{387--396}.
\newblock


\bibitem[\protect\citeauthoryear{Li, Hu, Jian, and Liu}{Li
  et~al\mbox{.}}{2016a}]%
        {li2016toward}
\bibfield{author}{\bibinfo{person}{Jundong Li}, \bibinfo{person}{Xia Hu},
  \bibinfo{person}{Ling Jian}, {and} \bibinfo{person}{Huan Liu}.}
  \bibinfo{year}{2016}\natexlab{a}.
\newblock \showarticletitle{Toward time-evolving feature selection on dynamic
  networks}. In \bibinfo{booktitle}{\emph{ICDM}}. IEEE,
  \bibinfo{pages}{1003--1008}.
\newblock


\bibitem[\protect\citeauthoryear{Li, Jia, Wang, Zhao, and Cheng}{Li
  et~al\mbox{.}}{2016b}]%
        {li2016predicting}
\bibfield{author}{\bibinfo{person}{Manling Li}, \bibinfo{person}{Yantao Jia},
  \bibinfo{person}{Yuanzhuo Wang}, \bibinfo{person}{Zeya Zhao}, {and}
  \bibinfo{person}{Xueqi Cheng}.} \bibinfo{year}{2016}\natexlab{b}.
\newblock \showarticletitle{Predicting Links and Their Building Time: A
  Path-Based Approach}. In \bibinfo{booktitle}{\emph{AAAI}}.
  \bibinfo{pages}{4228--4229}.
\newblock


\bibitem[\protect\citeauthoryear{Liben-Nowell and Kleinberg}{Liben-Nowell and
  Kleinberg}{2007}]%
        {liben2007link}
\bibfield{author}{\bibinfo{person}{David Liben-Nowell} {and}
  \bibinfo{person}{Jon Kleinberg}.} \bibinfo{year}{2007}\natexlab{}.
\newblock \showarticletitle{The link-prediction problem for social networks}.
\newblock \bibinfo{journal}{\emph{Journal of the American Society for
  Information Science and Technology}} \bibinfo{volume}{58},
  \bibinfo{number}{7} (\bibinfo{year}{2007}), \bibinfo{pages}{1019--1031}.
\newblock


\bibitem[\protect\citeauthoryear{McPherson, Smith-Lovin, and Cook}{McPherson
  et~al\mbox{.}}{2001}]%
        {mcpherson2001birds}
\bibfield{author}{\bibinfo{person}{Miller McPherson}, \bibinfo{person}{Lynn
  Smith-Lovin}, {and} \bibinfo{person}{James~M Cook}.}
  \bibinfo{year}{2001}\natexlab{}.
\newblock \showarticletitle{Birds of a feather: Homophily in social networks}.
\newblock \bibinfo{journal}{\emph{Annual Review of Sociology}}
  (\bibinfo{year}{2001}), \bibinfo{pages}{415--444}.
\newblock


\bibitem[\protect\citeauthoryear{Myers and Leskovec}{Myers and
  Leskovec}{2014}]%
        {myers2014bursty}
\bibfield{author}{\bibinfo{person}{Seth~A Myers} {and} \bibinfo{person}{Jure
  Leskovec}.} \bibinfo{year}{2014}\natexlab{}.
\newblock \showarticletitle{The bursty dynamics of the twitter information
  network}. In \bibinfo{booktitle}{\emph{WWW}}. ACM, \bibinfo{pages}{913--924}.
\newblock


\bibitem[\protect\citeauthoryear{Newman}{Newman}{2005}]%
        {newman2005power}
\bibfield{author}{\bibinfo{person}{Mark~EJ Newman}.}
  \bibinfo{year}{2005}\natexlab{}.
\newblock \showarticletitle{Power laws, Pareto distributions and Zipf's law}.
\newblock \bibinfo{journal}{\emph{Contemporary Physics}} \bibinfo{volume}{46},
  \bibinfo{number}{5} (\bibinfo{year}{2005}), \bibinfo{pages}{323--351}.
\newblock


\bibitem[\protect\citeauthoryear{Nguyen, Lim, Tan, Jiang, and Sun}{Nguyen
  et~al\mbox{.}}{2010}]%
        {nguyen2010you}
\bibfield{author}{\bibinfo{person}{Viet-An Nguyen}, \bibinfo{person}{Ee-Peng
  Lim}, \bibinfo{person}{Hwee-Hoon Tan}, \bibinfo{person}{Jing Jiang}, {and}
  \bibinfo{person}{Aixin Sun}.} \bibinfo{year}{2010}\natexlab{}.
\newblock \showarticletitle{Do You Trust to Get Trust? A Study of Trust
  Reciprocity Behaviors and Reciprocal Trust Prediction.}. In
  \bibinfo{booktitle}{\emph{SDM}}. SIAM, \bibinfo{pages}{72--83}.
\newblock


\bibitem[\protect\citeauthoryear{Papadopoulos, Kompatsiaris, Vakali, and
  Spyridonos}{Papadopoulos et~al\mbox{.}}{2012}]%
        {papadopoulos2012community}
\bibfield{author}{\bibinfo{person}{Symeon Papadopoulos},
  \bibinfo{person}{Yiannis Kompatsiaris}, \bibinfo{person}{Athena Vakali},
  {and} \bibinfo{person}{Ploutarchos Spyridonos}.}
  \bibinfo{year}{2012}\natexlab{}.
\newblock \showarticletitle{Community detection in social media}.
\newblock \bibinfo{journal}{\emph{Data Mining and Knowledge Discovery}}
  \bibinfo{volume}{24}, \bibinfo{number}{3} (\bibinfo{year}{2012}),
  \bibinfo{pages}{515--554}.
\newblock


\bibitem[\protect\citeauthoryear{Rodriguez, Leskovec, Balduzzi, and
  Sch{\"o}lkopf}{Rodriguez et~al\mbox{.}}{2014}]%
        {rodriguez2014uncovering}
\bibfield{author}{\bibinfo{person}{Manuel~Gomez Rodriguez},
  \bibinfo{person}{Jure Leskovec}, \bibinfo{person}{David Balduzzi}, {and}
  \bibinfo{person}{Bernhard Sch{\"o}lkopf}.} \bibinfo{year}{2014}\natexlab{}.
\newblock \showarticletitle{Uncovering the structure and temporal dynamics of
  information propagation}.
\newblock \bibinfo{journal}{\emph{Network Science}} \bibinfo{volume}{2},
  \bibinfo{number}{01} (\bibinfo{year}{2014}), \bibinfo{pages}{26--65}.
\newblock


\bibitem[\protect\citeauthoryear{Sarkar, Chakrabarti, and Jordan}{Sarkar
  et~al\mbox{.}}{2012}]%
        {sarkar2012nonparametric}
\bibfield{author}{\bibinfo{person}{Purnamrita Sarkar},
  \bibinfo{person}{Deepayan Chakrabarti}, {and} \bibinfo{person}{Michael~I
  Jordan}.} \bibinfo{year}{2012}\natexlab{}.
\newblock \showarticletitle{Nonparametric Link Prediction in Dynamic Networks}.
  In \bibinfo{booktitle}{\emph{ICML}}. \bibinfo{pages}{1687--1694}.
\newblock


\bibitem[\protect\citeauthoryear{Squartini, Picciolo, Ruzzenenti, and
  Garlaschelli}{Squartini et~al\mbox{.}}{2013}]%
        {squartini2013reciprocity}
\bibfield{author}{\bibinfo{person}{Tiziano Squartini},
  \bibinfo{person}{Francesco Picciolo}, \bibinfo{person}{Franco Ruzzenenti},
  {and} \bibinfo{person}{Diego Garlaschelli}.} \bibinfo{year}{2013}\natexlab{}.
\newblock \showarticletitle{Reciprocity of weighted networks}.
\newblock \bibinfo{journal}{\emph{Scientific Reports}}  \bibinfo{volume}{3}
  (\bibinfo{year}{2013}).
\newblock


\bibitem[\protect\citeauthoryear{Sun, Han, Aggarwal, and Chawla}{Sun
  et~al\mbox{.}}{2012}]%
        {sun2012will}
\bibfield{author}{\bibinfo{person}{Yizhou Sun}, \bibinfo{person}{Jiawei Han},
  \bibinfo{person}{Charu~C Aggarwal}, {and} \bibinfo{person}{Nitesh~V Chawla}.}
  \bibinfo{year}{2012}\natexlab{}.
\newblock \showarticletitle{When will it happen?: relationship prediction in
  heterogeneous information networks}. In \bibinfo{booktitle}{\emph{WSDM}}.
  ACM, \bibinfo{pages}{663--672}.
\newblock


\bibitem[\protect\citeauthoryear{Szell, Lambiotte, and Thurner}{Szell
  et~al\mbox{.}}{2010}]%
        {szell2010multirelational}
\bibfield{author}{\bibinfo{person}{Michael Szell}, \bibinfo{person}{Renaud
  Lambiotte}, {and} \bibinfo{person}{Stefan Thurner}.}
  \bibinfo{year}{2010}\natexlab{}.
\newblock \showarticletitle{Multirelational organization of large-scale social
  networks in an online world}.
\newblock \bibinfo{journal}{\emph{PNAS}} \bibinfo{volume}{107},
  \bibinfo{number}{31} (\bibinfo{year}{2010}), \bibinfo{pages}{13636--13641}.
\newblock


\bibitem[\protect\citeauthoryear{Tylenda, Angelova, and Bedathur}{Tylenda
  et~al\mbox{.}}{2009}]%
        {tylenda2009towards}
\bibfield{author}{\bibinfo{person}{Tomasz Tylenda}, \bibinfo{person}{Ralitsa
  Angelova}, {and} \bibinfo{person}{Srikanta Bedathur}.}
  \bibinfo{year}{2009}\natexlab{}.
\newblock \showarticletitle{Towards time-aware link prediction in evolving
  social networks}. In \bibinfo{booktitle}{\emph{SNA-KDD}}. ACM.
\newblock


\bibitem[\protect\citeauthoryear{Wang, Pedreschi, Song, Giannotti, and
  Barabasi}{Wang et~al\mbox{.}}{2011}]%
        {wang2011human}
\bibfield{author}{\bibinfo{person}{Dashun Wang}, \bibinfo{person}{Dino
  Pedreschi}, \bibinfo{person}{Chaoming Song}, \bibinfo{person}{Fosca
  Giannotti}, {and} \bibinfo{person}{Albert-Laszlo Barabasi}.}
  \bibinfo{year}{2011}\natexlab{}.
\newblock \showarticletitle{Human mobility, social ties, and link prediction}.
  In \bibinfo{booktitle}{\emph{SIGKDD}}. ACM, \bibinfo{pages}{1100--1108}.
\newblock


\bibitem[\protect\citeauthoryear{Weng, Lim, Jiang, and He}{Weng
  et~al\mbox{.}}{2010}]%
        {weng2010twitterrank}
\bibfield{author}{\bibinfo{person}{Jianshu Weng}, \bibinfo{person}{Ee-Peng
  Lim}, \bibinfo{person}{Jing Jiang}, {and} \bibinfo{person}{Qi He}.}
  \bibinfo{year}{2010}\natexlab{}.
\newblock \showarticletitle{Twitterrank: finding topic-sensitive influential
  twitterers}. In \bibinfo{booktitle}{\emph{WSDM}}. ACM,
  \bibinfo{pages}{261--270}.
\newblock


\bibitem[\protect\citeauthoryear{Wu and Carroll}{Wu and Carroll}{1988}]%
        {wu1988estimation}
\bibfield{author}{\bibinfo{person}{Margaret~C Wu} {and}
  \bibinfo{person}{Raymond~J Carroll}.} \bibinfo{year}{1988}\natexlab{}.
\newblock \showarticletitle{Estimation and comparison of changes in the
  presence of informative right censoring by modeling the censoring process}.
\newblock \bibinfo{journal}{\emph{Biometrics}} (\bibinfo{year}{1988}),
  \bibinfo{pages}{175--188}.
\newblock


\bibitem[\protect\citeauthoryear{Yu, Cheng, Aggarwal, Chen, and Wang}{Yu
  et~al\mbox{.}}{2017}]%
        {yu2017link}
\bibfield{author}{\bibinfo{person}{Wenchao Yu}, \bibinfo{person}{Wei Cheng},
  \bibinfo{person}{Charu~C Aggarwal}, \bibinfo{person}{Haifeng Chen}, {and}
  \bibinfo{person}{Wei Wang}.} \bibinfo{year}{2017}\natexlab{}.
\newblock \showarticletitle{Link prediction with spatial and temporal
  consistency in dynamic networks}. In \bibinfo{booktitle}{\emph{IJCAI}}.
  \bibinfo{pages}{3343--3349}.
\newblock


\bibitem[\protect\citeauthoryear{Zhu, Guo, Yin, Ver~Steeg, and Galstyan}{Zhu
  et~al\mbox{.}}{2016}]%
        {zhu2016scalable}
\bibfield{author}{\bibinfo{person}{Linhong Zhu}, \bibinfo{person}{Dong Guo},
  \bibinfo{person}{Junming Yin}, \bibinfo{person}{Greg Ver~Steeg}, {and}
  \bibinfo{person}{Aram Galstyan}.} \bibinfo{year}{2016}\natexlab{}.
\newblock \showarticletitle{Scalable temporal latent space inference for link
  prediction in dynamic social networks}.
\newblock \bibinfo{journal}{\emph{IEEE Transactions on Knowledge and Data
  Engineering}} \bibinfo{volume}{28}, \bibinfo{number}{10}
  (\bibinfo{year}{2016}), \bibinfo{pages}{2765--2777}.
\newblock


\end{thebibliography}
\end{document}